\documentclass[prb,preprint,amsmath,amssymb,superscriptaddress,notitlepage]{revtex4-1}
\usepackage{epsfig}
\usepackage{graphicx}
\usepackage{color}
\usepackage{bm}
\usepackage{fixme}
\usepackage{braket}

\newcommand{\rb}{\mathbf{r}}
\newcommand{\kb}{\mathbf{k}}
\newcommand{\calE}{{\epsilon}}

\begin{document}

\title{Unidirectional magnetoresistance and spin-orbit torque in NiMnSb}

\author{J.~\v{Z}elezn\'y} 
\affiliation{Institute of Physics, Czech Academy of Sciences, Cukrovarnick\'a 10, 162 00, Praha 6, Czech Republic}
\author{Z.~Fang} 
\affiliation{Cavendish Laboratory, University of Cambridge, CB3 0HE, United Kingdom}
\author{K.~Olejn\'ik} 
\affiliation{Institute of Physics, Czech Academy of Sciences, Cukrovarnick\'a 10, 162 00, Praha 6, Czech Republic}
\author{J.~Patchett} 
\affiliation{Cavendish Laboratory, University of Cambridge, CB3 0HE, United Kingdom}
\author{F.~Gerhard}
\affiliation{Physikalisches Institut (EP3), Universit\"{a}t W\"{u}rzburg, Am Hubland, D-97074 W\"{u}rzburg, Germany}
\author{C.~Gould}
\affiliation{Physikalisches Institut (EP3), Universit\"{a}t W\"{u}rzburg, Am Hubland, D-97074 W\"{u}rzburg, Germany}
\author{L.~W. Molenkamp}
\affiliation{Physikalisches Institut (EP3), Universit\"{a}t W\"{u}rzburg, Am Hubland, D-97074 W\"{u}rzburg, Germany}
\author{C.~Gomez-Olivella}
\affiliation{Institute of Physics, Czech Academy of Sciences, Cukrovarnick\'a 10, 162 00, Praha 6, Czech Republic}
\author{J.~Zemen}
\affiliation{Faculty of Electrical Engineering, Czech Technical University in Prague, Technick\'a 2, Prague 166 27, Czech Republic}
\author{T.~Tich\'y}
\affiliation{Faculty of Electrical Engineering, Czech Technical University in Prague, Technick\'a 2, Prague 166 27, Czech Republic}
\author{T.~Jungwirth}
\affiliation{Institute of Physics, Czech Academy of Sciences, Cukrovarnick\'a 10, 162 00, Praha 6, Czech Republic} 
\affiliation{School of Physics and Astronomy, University of Nottingham, Nottingham NG7 2RD, United Kingdom}
\author{C.~Ciccarelli}
\affiliation{Cavendish Laboratory, University of Cambridge, CB3 0HE, United Kingdom}

\begin{abstract}
Spin-dependent transport phenomena due to relativistic spin-orbit coupling and broken space-inversion symmetry are often difficult to interpret microscopically, in particular when occurring at surfaces or interfaces. Here we present a theoretical and experimental study of spin-orbit torque and unidirectional magnetoresistance in a model room-temperature ferromagnet NiMnSb with inversion asymmetry in the bulk of this half-heusler crystal. Besides the angular dependence on magnetization, the competition of Rashba and Dresselhaus-like spin-orbit couplings results in the dependence of these effects on the crystal direction of the applied electric field. The phenomenology that we observe highlights potential inapplicability of commonly considered approaches for interpreting experiments.  We point out that, in general, there is no direct link between the current-induced non-equilibrium spin polarization inferred from the measured spin-orbit torque and the unidirectional magnetiresistance. We also emphasize that the unidirectional magnetoresistance has not only longitudinal but also transverse components in the electric field -- current indices which complicates its separation from the thermoelectric contributions to the detected signals in common experimental techniques. We use the theoretical results to analyze our measurements of the on-resonance and off-resonance mixing signals in microbar devices fabricated from an epitaxial NiMnSb film along different crystal directions. Based on the analysis we extract an experimental estimate of the unidirectional magnetoresistance in NiMnSb.
\end{abstract}

\maketitle

\section{Introduction}

Anisotropic magnetoresistance (AMR) is an example of a relativistic transport phenomenon in ferromagnets with a long history. McGuire and Potter\cite{McGuire1975} provided a phenomenological model that fully explained the angular dependence of AMR. The model was based on a general argument that the magnetisation dependent conductivity tensor $\sigma_{ij}(\bf m)$, like any other observable,\cite{Neumann1885} reflects the symmetry of the ferromagnetic crystal. This means that if a symmetry operation belongs to the crystallographic point group, the conductivity tensor is left invariant under the same symmetry operation. As an example, when an electrical  current flows in an isotropic (polycrystalline)  ferromagnet, the magnetisation direction defines the only axis of rotational symmetry, which results in a $\cos(2\theta)$ angular dependence of AMR, with $\theta$ being the angle between the magnetisation and the current.

While AMR can be observed in any magnetic system, ferromagnetic conductors with an inversion asymmetric crystal structure have been more recently identified as a fruitful platform for discovering and utilizing a range of new relativistic spintronics phenomena.\cite{Bernevig2005a,Chernyshov2009,Fang2011,Kurebayashi2014,Ciccarelli2014,Ciccarelli2016} Apart from the exchange field ${\bf B}_{\rm ex}$, which splits the electronic structure in majority and minority spin bands, the inversion symmetry breaking in the lattice together with spin-orbit coupling introduces an additional splitting $\Delta H_{\rm SO}={\bf B}_{\rm SO}({\bf k})\cdot{\bf s}$. Here ${\bf B}_{\rm SO}({\bf k})$ is an effective magnetic field that depends on the crystal momentum ${\bf k}$, while ${\bf s}$ represents the spin-polarisation. The key property of  ${\bf B}_{\rm SO}({\bf k})$ is that it is odd in ${\bf k}$ and thus it results in opposite spin splitting for opposite ${\bf k}$. In inversion asymmetric strained zinc-blende semiconductors like GaAs,  ${\bf B}_{\rm SO}({\bf k})$ is a combination of Rashba and Dresselhaus symmetry terms, ${\bf B}^{\rm R}_{\rm SO}({\bf k})\propto(k_y,-k_x)$ and ${\bf B}^{\rm D}_{\rm SO}({\bf k})\propto(k_x,-k_y)$. A direct manifestation of this splitting is found in the spin-orbit torque (SOT) \cite{Manchon2019} in, e.g.,  (Ga,Mn)As that emerge when an electrical current is applied to this inversion-asymmetric ferromagnetic semiconductor.\cite{Bernevig2005a,Chernyshov2009,Fang2011,Kurebayashi2014} In metallic systems such as half-heusler NiMnSb studied here, ${\bf B}_{\rm SO}({\bf k})$ is not described by the simple linear-in-${\bf k}$ Rashba and Dresselhaus form. SOT, nevertheless, contains Rashba and Dreselhaus-like terms analogous to those observed in the zinc-blende semiconductors as those reflect the crystal symmetry which is the same in zinc-blende and half-heusler crystals.\cite{Ciccarelli2016} 

SOT in bulk non-centrosymmetric crystals is associated with a current-induced non-equilibrium  spin polarisation $\delta {\bf s}_{\rm SO}$, an effect often referred to as the Edelstein effect or the inverse spin-galvanic effect (iSGE).\cite{Silov2004,Kato2004b,Ganichev2004a,Wunderlich2005,Sinova2015,Manchon2019} When $\delta {\bf s}_{\rm SO}$ is perpendicular to the magnetization, it will exert a torque on it via the exchange coupling. The component of $\delta {\bf s}_{\rm SO}$ that is parallel to  ${\bf B}_{\rm ex}$ does not lead to a magnetization torque and is, therefore, transparent to any experimental method that relies on driving the magnetisation out of equilibrium by SOT.\cite{Fang2011,Garello2013,Manchon2019} 

Besides SOT, ${\bf B}_{\rm SO}({\bf k})$ together with  ${\bf B}_{\rm ex}$ can also lead to magneto-transport terms that are second order in the applied electric field and, unlike the first-order AMR, are odd under the magnetization. The unidirectional magnetoresistance (UMR) is an example of such a second-order magneto-transport effect that  was previously reported in non-centrosymmetric ferromagnet/paramagnet bilayers or bulk ferromagnets.\cite{Avci2015,Olejnik2015} The origin of UMR was considered to be closely connected to the phenomenology of the giant magnetoresistance (GMR).\cite{Baibich1988,Binasch1989,Olejnik2015,Avci2018} While in the GMR multilayer, the fixed reference ferromagnetic layer acts as an external source of spin $\delta {\bf s}$, in UMR this is replaced with $\delta {\bf s}_{\rm SO}$  generated internally by the spin-orbit coupling. As for GMR, spin-dependent scattering within the ferromagnet results in this scenario in a different resistance depending on the orientation of $\delta {\bf s}_{\rm SO}$ relative  to the magnetization of the probed ferromagnet. Moreover, the accumulated spin can introduce a proportional change in the exchange splitting of the bands, further affecting the conductivity by influencing spin transmission of minority and majority spins. Another UMR mechanism considers that the magnons' population of the ferromagnet is increased or decreased depending on the orientation of $\delta {\bf s}_{\rm SO}$ relative to the magnetisation. This leads to a change in the average magnetisation, which also results in a net change of the magnetoresistance. Although it is possible to experimentally distinguish these different contributions for their  particular dependence on current density and magnetic field,\cite{Avci2018,Langenfeld2016} they all share a common origin in the component of $\delta {\bf s}_{\rm SO}$ collinear with the ferromagnet's magnetisation. 

In Sec.~II we report our symmetry analysis and {\em ab initio} calculations of current induced spin polarization and UMR in NiMnSb. We use this model system to highlight potential misconceptions when inferring these quantities from experiment. In Sec.~III we then discuss our measurements in NiMnSb microbars patterned along different crystal directions, and in Sec.~IV we summarize our main results.

\section{Theoretical results and general implications for the analysis  of experiments}
\subsection{Current-induced spin polarization and spin-orbit torque}
The tetragonal distortion of the non-centrosymmetric cubic unit cell of NiMnSb, induced by the lattice mismatch with the substrate, lowers its symmetry to a $-42m$ symmetry point group and results in a Dresselhaus-like $k$-linear term of the spin-orbit coupling. Experiments in NiMnSb epilayers show an additional Rashba-like $k$-linear term of the spin-orbit coupling which we model by introducing a shear strain.\cite{Ciccarelli2016} This lowers the symmetry further to a point group $mm2$. When an electrical current is passed in the plane perpendicular to the growth direction, carriers acquire a non-equilibrium spin polarization $\delta {\bf s}_{\rm SO}$, which in general can be decomposed into a component that is parallel to the in-plane magnetisation of the NiMnSb film, $\delta {\bf s}^{\parallel}_{\rm SO}$, and a component that is perpendicular to it, $\delta {\bf s}^{\perp}_{\rm SO}$. This includes both intrinsic and extrinsic contributions.

We  use {\em ab-initio} calculations to evaluate the current-induced non-equilibrium spin polarization as a function of the  directions of magnetization and applied electric field (see Appendix \ref{appendix:calculations} for the description of the numerical method). We consider here only the in-plane components of current-induced spin-polarization since this is the component relevant for the UMR and the field-like torque. For these calculations we used the relaxation time $\tau$ chosen so that the theoretical conductivity matches the experimental value of $3.3\times10^4$~Scm$^{-1}$. The tetragonal and shear strains are chosen to make the Rashba and Dresselhaus contributions of comparable strength,\cite{Ciccarelli2016} resulting in a non-trivial dependence of $\delta {\bf s}_{\rm SO}$ on the crystal direction of the applied electric field. In Fig.~1 we plot the perpendicular ($\delta {\bf s}^{\perp}_{\rm SO}$) and parallel ($\delta {\bf s}^{\parallel}_{\rm SO}$) components of the current-induced non-equilibrium spin polarization. The dependence of  $\delta {\bf s}^{\perp}_{\rm SO}\sim \cos(\theta-\theta_{\rm SO})$ on the magnetization angle $\theta$ corresponds to what we would expect to find for a field-like SOT generated by a magnetisation-independent effective spin-orbit field, ${\bf h}_{\rm SO}\approx -J_{\rm ex}\delta {\bf s}_{\rm SO}/m$, where $J_{\rm ex}$ is the exchange constant between carrier spins and the 
in-plane magnetisation {\rm\bf m}. $\theta_{\rm SO}$ would then correspond to the angle of $\delta {\bf s}_{\rm SO}$ for the given crystal direction of the applied electric field. In our case, $\theta_{\rm SO}=0$ for the electric field along $[110]$ or $[1\bar{1}0]$ axes since the Rashba and Dresselhaus-like spin-orbit fields are (anti)parallel for these two crystal directions. For $[100]$ or $[010]$ axes, the two spin-orbit fields are orthogonal to each other and their vector sum results in  $\theta_{\rm SO}\neq 0$.

Remarkably, when we include  $\delta {\bf s}^{\parallel}_{\rm SO}$, total $\delta {\bf s}_{\rm SO}$  becomes magnetization-dependent (see Figs.~1). We emphasize that this dependence of $\delta {\bf s}_{\rm SO}$ (${\bf h}_{\rm SO}$) on the magnetization angle would be invisible when measuring SOT, which only depends on $\delta {\bf s}^{\perp}_{\rm SO}$. SOT  can be reliably obtained from on-resonance mixing-signal measurements since the detected signal can be directly attributed to the precessing magnetization driven by the SOT. Our calculations  in Fig.~1  demonstrate, however, that a simple harmonic dependence of $\delta {\bf s}^{\perp}_{\rm SO}$ on the magnetization angle, when inferred from the SOT measurement, does not imply that the total $\delta {\bf s}_{\rm SO}$ is constant and that its parallel component $\delta {\bf s}^{\parallel}_{\rm SO}$ is a 90$^{\circ}$-phase shifted replica of the perpendicular component. Therefore, if considering $\delta {\bf s}^{\parallel}_{\rm SO}$ as the driving mechanism behind UMR, SOT measurements cannot be, in general, used to quantify experimentally $\delta {\bf s}^{\parallel}_{\rm SO}$ in a given structure. 


\subsection{Unidirectional magnetoresistance}

When writing the total current density $j$ up to the second order in the applied electric filed $E$ as,
\begin{equation}
\label{UMR}
j_i=j_i^{(1)}+j_i^{(2)}=\sigma_{ij}E_j+\xi_{ijk}E_jE_k\, ,
\end{equation}
UMR has been associated with the longitudinal component of the $\xi_{ijk}$  transport coefficient.\cite{Avci2015,Olejnik2015} Formally, $\xi_{ijk}$ is described by the second order quantum mechanical Kubo formula.  However, finding accurate solutions of the formula is a major challenge, in particular in the presence of electron scattering. Here we analyze $\xi_{ijk}$ using a semiclassical Boltzmann approximation, where 
\begin{equation}
\label{Boltzmann}
\xi_{ijk}=-\frac{e^3\tau^2}{2}\sum_n\int\frac{d{\bf k}}{(2\pi)^3} v_n^i v_n^j v_n^k \frac{\partial^2f_0}{\partial\epsilon_n({\bf k})^2}\, ,
\end{equation}
and ${\bf v}_n({\bf k})=\frac{1}{\hbar}\frac{\partial\epsilon_n}{\partial{\bf k}}$, $f_0$ is the Fermi-Dirac equilibrium distribution function, and $\epsilon_n({\bf k})$ is the band energy (see Appendix \ref{appendix:derivation} for the derivation of this formula).

Since the group velocity ${\bf v}({\bf k})$ is odd under space inversion, the second-order term will vanish in crystals that have inversion symmetry. Moreover,  it will also vanish in nonmagnetic crystals since ${\bf v}({\bf k})$ is odd under time-reversal. Furthermore, similarly to the anomalous Hall effect in coplanar magnetic systems, the second-order term will vanish in the absence of spin-orbit coupling, as the system will then be invariant under combined spin rotation and time-reversal symmetry.\cite{Zhang2018h} $\xi_{ijk}$ will thus be present in magnetic crystals with broken inversion symmetry as a consequence of the spin-orbit coupling. These are the same symmetry requirements as for the existence of SOT. 

In Fig.~2 we plot the calculations of the longitudinal and transverse components of the second-order current obtained from the Boltzmann equation (\ref{Boltzmann}), normalized to the first-order current as a function of the magnetization angle for different directions of the electric field and for a current density of $10^{10}$~Am$^{-2}$. The relative  amplitudes of the longitudinal component for different directions of the applied electric field are similar to the relative amplitudes of $\delta {\bf s}_{\rm SO}^\perp$ (cf. Figs.~1 and 2). In both cases, the amplitudes are comparable for the electric field along $[100]$ and $[010]$ axes, while the largest/smallest amplitude is obtained for fields along $[110]/[1\bar{1}0]$ axes. From these results we can expect similar amplitude ratios also in the measured  UMR and SOT. We point out, however, that this is not necessarily a consequence of a common proportionality of UMR and SOT to $\delta {\bf s}_{\rm SO}$. The Boltzmann approximation formula (\ref{Boltzmann})  gives an explicit example of a contribution to UMR with no direct relationship to $\delta {\bf s}_{\rm SO}$. The similar amplitude ratios in the two cases are merely a reflection of a common Rashba-Dresselhaus-like symmetry of the underlying spin-orbit coupled electronic structure. 

Our calculations in Fig.~2 also illustrate that the second-order current can have a sizable transverse component. Specifically in NiMnSb, the transverse component has a comparable amplitude to the longitudinal component for $[100]$ and $[010]$ crystal directions. Here we recall that in earlier experimental studies, the separation of UMR from competing thermo-electric contributions was based on the assumption that UMR had only a longitudinal component while the thermal effects contributed to both longitudinal and transverse signals.\cite{Avci2015,Olejnik2015} Our results show that using the transverse signal for experimentally calibrating the magnitude of the thermal contribution is not, in general, reliable because the transverse component can also contain a contribution from UMR.


\section{Measured data and discussion}

In our experiments, we pattern all our bard from the same 34~nm thick film of ferromagnetic NiMnSb epitaxially grown on a 200~nm thick In$_{0.53}$Ga$_{0.47}$As buffer layer on an Fe:InP insulating substrate and capped with a 5~nm MgO layer. The vertical lattice constant of 5.97 \AA~ indicates that the film is under compressive strain and is close to a stoichiometric composition.\cite{Gerhard2014a}  Fig.~3a illustrates our measurement set-up. The NiMnSb film is patterned into $4\times40$~$\mu$m$^2$ bars along different crystal directions and mounted on the rotational stage of an electromagnet. A microwave current $I_0\cos(\omega t)$ is passed in the bars and the polarizing action of the spin-orbit coupling induces an oscillating non-equilibrium spin population $\delta {\bf s}_{\rm SO}(\omega)$ which scales linearly with the current. The transverse component $\delta {\bf s}^{\perp}_{\rm SO}$ is responsible for generating torques on the magnetisation via the effective field ${\bf h}^{\perp}_{\rm SO}\approx -J_{\rm ex}\delta {\bf s}^{\perp}_{\rm SO}/m$. As described in the previous paragraph, the longitudinal component $\delta{S}^{||}$ is responsible for generating both longitudinal and transverse components of the unidirectional magnetoresistance. Here we focus on the longitudinal components only and measure the dc longitudinal voltage $V_{\rm dc}$ via a bias tee.

When ferromagnetic resonance is excited, rectification between the microwave current and the oscillating AMR results in a resonance in $V_{\rm dc}$.\cite{Fang2011} The resonance is clearly visible in the 2D plots in Fig.~3b. It shows $V_{\rm dc}$ as a function of the external magnetic field and its direction $\theta$ with respect to the current direction, for  bars parallel to the [$\bar{1}$10] and [100] axes. From the analysis of its line-shape (see Ref.~\onlinecite{Kurebayashi2014} for details) we are able to quantify ${\bf h}^{\perp}_{\rm SO}$. In agreement with previous studies on different systems,\cite{Fang2011,Kurebayashi2014,Garello2013} we experimentally identify the $\theta$-dependence of ${\bf h}^{\perp}_{{\rm SO}}$ for  the different crystal directions in which the microbars are patterned, as shown in Fig.~4. As in the theoretical calculations of $\delta {\bf s}^{\perp}_{\rm SO}$, the amplitudes of measured ${\bf h}^{\perp}_{\rm SO}$ for bars along the [100] and [010] axes are similar, while the largest/smallest amplitudes are obtained for the $[110]/[1\bar{1}0]$ directions, consistent with the combined Rashba-Dresselhaus-like symmetry of the NiMnSb electronic struture. The theoretical  magnitude of $|{\bf h}^{\perp}_{\rm SO}|\approx |J_{\rm ex}\delta {\bf s}^{\perp}_{\rm SO}/m|\sim10$~$\mu$T at $10^{10}$~Am$^{-2}$ current density was reported earlier in Ref.~\onlinecite{Ciccarelli2016} and is of the same order of magnitude as in the experiment.

In Fig. 3b we notice that the resonance is sitting on a sinusoidally varying background, which we refer to as $V_{\rm BG}$. This is also shown in Fig.~5, where the $\theta$-dependence of the background voltage at a saturating magnetic field is plotted for  bars patterned along the four different crystal directions [100], [010], [110] and [$\bar{1}$10]. The striking feature is that the magnitude of $V_{\rm BG}$ is again strongly dependent on the bar direction, however, the directions corresponding to the largest and smallest $V_{\rm BG}$ switched place compared to ${\bf h}_{\rm SO}$ (or $\delta {\bf s}_{\rm SO}$). We note that these amplitude ratios of  $V_{\rm BG}$, as well as the crystal direction dependent phase shifts, are reproducible in different physical sets of samples patterned along the four crystal directions and do not depend on applied power, as shown in Appendix \ref{appendix:power}.

To interpret the measured $V_{\rm BG}$ we now consider UMR and the thermoelectric contribution, namely the anomalous Nernst effect (ANE). When exciting the system by the applied ac current,  an out-of-plane temperature gradient due to Joule heating can result in an electrical signal detected in the sample plane due to ANE.  Since the heat deposited by the current scales with the square of the current density, ANE is a second-order effect in the electric field, just like the second order term of the conductivity. Based on a recent experimental measurement\cite{Sharma2019} of ANE in NiMnSb and our numerical simulation of the heat gradient, we estimate that the contribution to $V_{\rm BG}$ due to ANE is $\sim 0.01-0.1$~$\mu$V per current density of $10^{10}$~Am$^{-2}$ (see Appendix \ref{appendix:ANE_modelling}). While the magnitude is similar to that of the measured signals in Fig.5, we do not expect a strong dependence of the ANE contribution to $V_{\rm BG}$ on the crystal direction of the applied current. This is because ANE requires only the time-reversal symmetry breaking by the magnetization while broken spatial symmetries of the crystal  only lead to additional, higher order corrections.

UMR, on the other hand, is generated by the inversion symmetry breaking which is of the combined Rashba-Dresselhaus-like form in our NiMnSb samples. As discussed in the theory section, this leads to a strong crystal direction dependence of the UMR. Since theory suggests that the amplitude ratios for the different crystal directions of SOT and UMR are similar, we can use this as a constraint when fitting the measured $V_{\rm BG}$ data. The results of the fitting shown in Fig. 5 were obtained by assuming fixed amplitude ratios of UMR, corresponding to the measured amplitude ratios of the  SOT fields ${\bf h}^{\perp}_{\rm SO}$, plus an ANE contribution with an amplitude which is independent of the crystal direction. The extracted  experimental ANE component is of the same order of magnitude as the above estimate. The fitted UMR  contribution is an order of magnitude larger than our theoretical value which we attribute to the crude semiclassical Boltzmann approximation used in the UMR calculations. In the future, more elaborate Kubo formula calculations seem necessary to capture UMR in NiMnSb on the quantitative level. They will also allow for verifying the correspondence between the SOT and UMR amplitude ratios, and by this for more firmly establishing the fitting method we used to separate the UMR and thermoelectric contributions in the measured data.

\section{Summary} 
Based on our study of ferromagnetic NiMnSb with non-centrosymmetric bulk crystal structure we make the following observations regarding the explored non-equilibrium spin-orbit coupling effects: (i) A harmonic dependence on the magnetization angle of the component of the current-induced spin polarization transverse to the magnetization does not imply, in general, that the in-plane component is its 90$^\circ$ phase shifted replica with the same amplitude. Despite the harmonic dependence of the transverse component, the  total spin polarization vector, and the corresponding total current-induced spin-orbit field vector, is not necessarily independent of the magnetization angle. As a result, a measurement of SOT, driven by the transverse component of the current-induced spin-orbit field, should not be used, in general, for extracting the total non-equilibrium spin-orbit field (spin polarization) vector. (ii) The approximate Boltzmann theory of UMR, together with the approximate assumption of SOT being proportional to the current-induced spin polarization, suggest that the amplitude ratios of UMR and SOT for electric fields applied along different crystal directions are similar. This can be used for separating experimental UMR and thermoelectric (e.g. ANE) contributions, by employing independently  measured SOT. On the other hand, measurements of the diagonal and off-diagonal components in the electric field -- current indices is not, in general, a reliable tool for separating UMR and thermoelectric contributions because both can have sizable diagonal and off-diagonal components. (iii) Finally, the Boltzmann theory also illustrates that UMR can have microscopic contributions which are not directly related to the current-induced spin polarization vector. Similar phenomenologies observed in UMR and SOT can be a mere reflection of the common underlying relativistic electronic structure with broken time and space inversion symmetries.
\begin{acknowledgments}
We  acknowledge  the  Grant  Agency  of  the  Czech  Republic  Grant  No.  19-18623Y, Ministry of Education of the
Czech Republic Grants LM2018110, LNSM-LNSpin and e-Infrastructure CZ – LM2018140, EU FET Open RIA Grant No. 766566  and  support  from  the  Institute  of Physics  of  the  Czech  Academy  of  Sciences  and  the  Max Planck Society through the Max Planck Partner Group Programme. CC acknowledges support from the Royal Society. JPP acknowledges support from the EPSRC. The work of JZ was supported by the Ministry of Education, Youth and Sports of the Czech Republic from the  OP RDE programme under the project International Mobility of Researchers MSCA-IF at CTU No. CZ.02.2.69/0.0/0.0/18\_070/0010457.
\end{acknowledgments}
%

\newpage
\begin{figure}[h]
\hspace*{-0cm}\includegraphics[width=.9\textwidth]{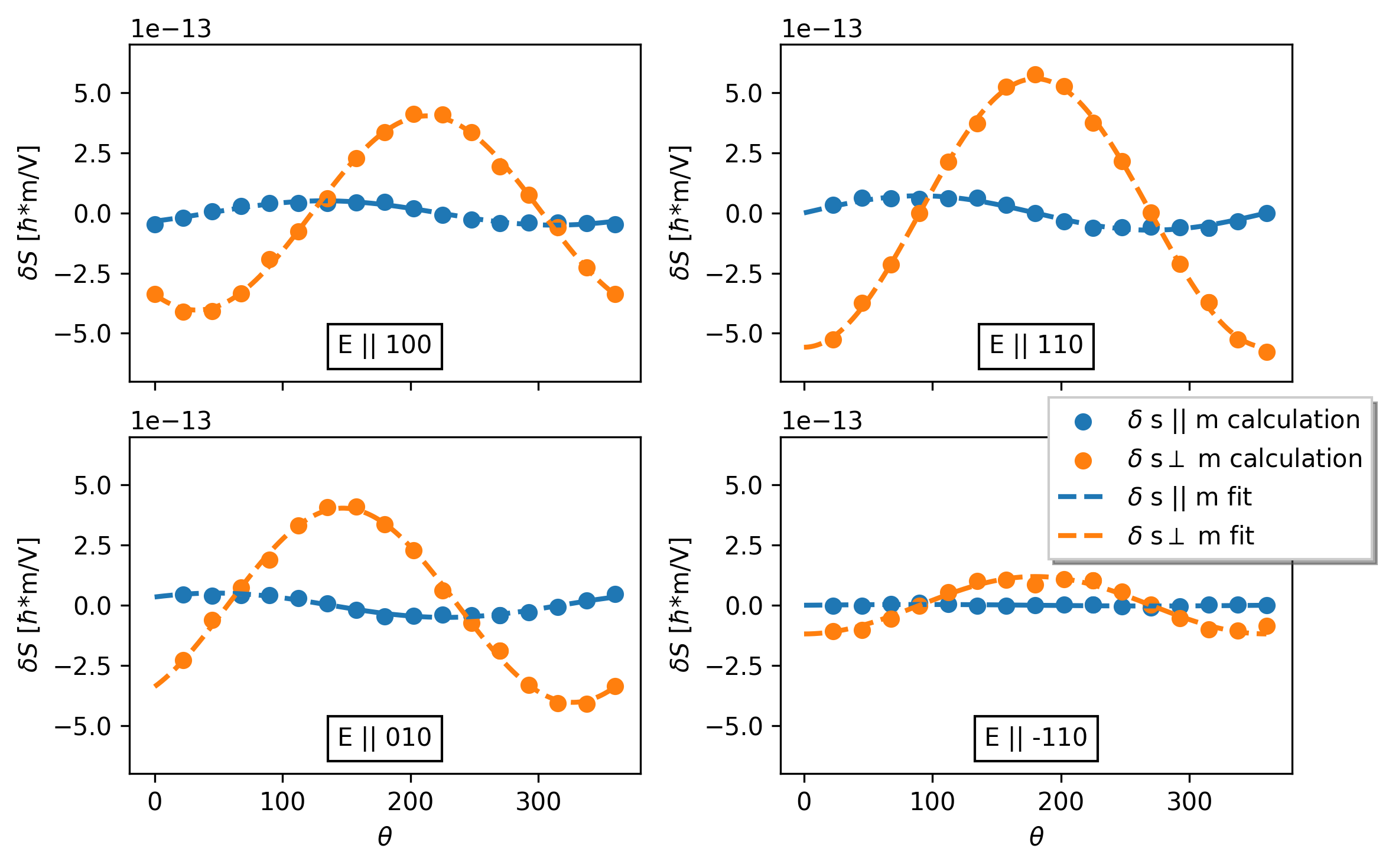}
\caption
{{\bf Calculations of current-induced spin polarization for different directions of electric field.} The plots show the total spin polarization, as well as its components that are longitudinal and transverse to the magnetization, as a function of the magnetization orientation within the (001)-plane. In each plot, $\theta$ denotes the angle of the magnetization measured from the electric field. Dashed lines represent harmonic fits to the numerical points.
}
\label{fig1}
\end{figure}

\begin{figure}[h]
\hspace*{-0cm}\includegraphics[width=.9\textwidth]{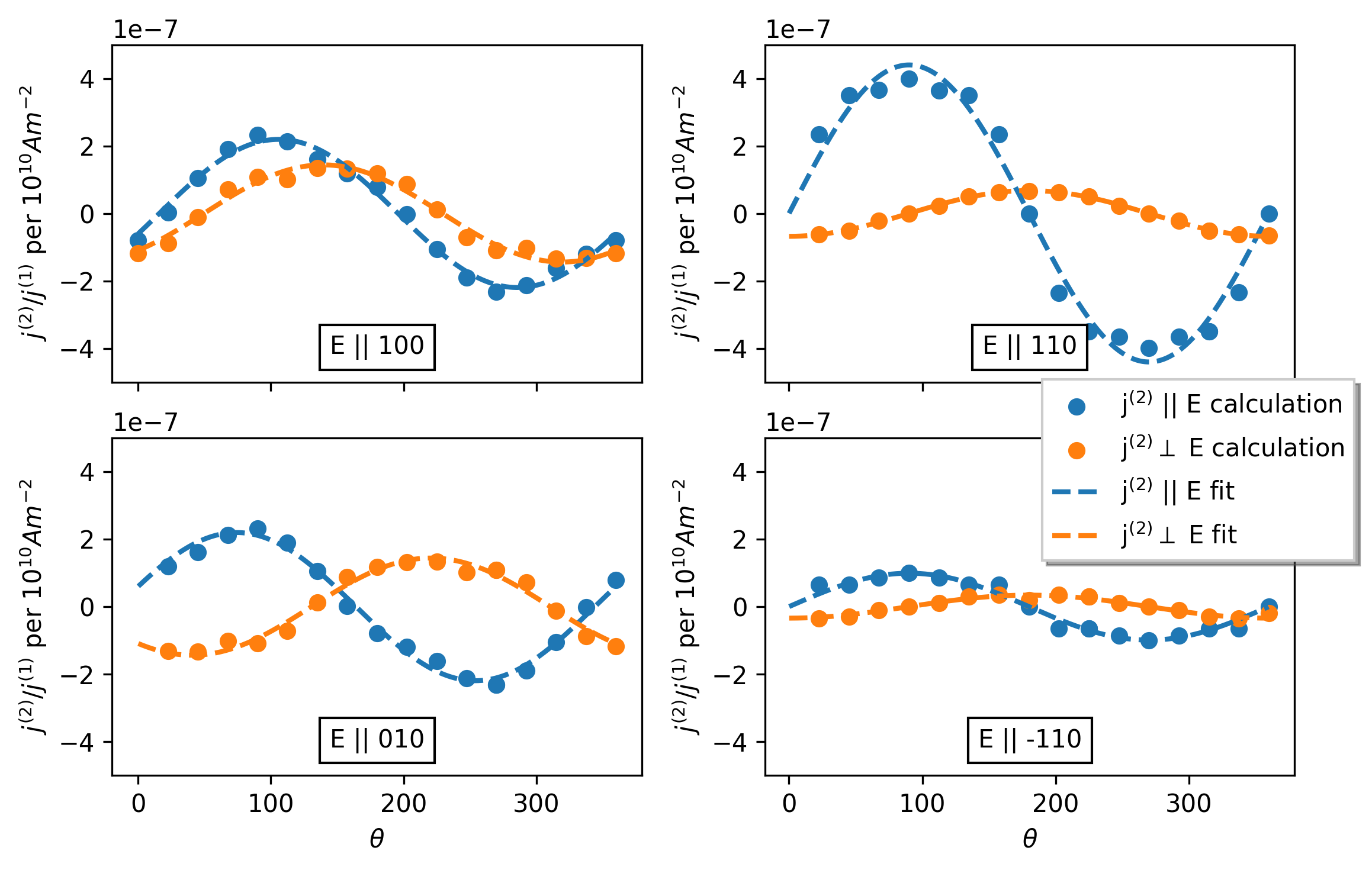}
\caption
{{\bf Second-order current calculations for different directions of the electric field.} The plots show the second order current separated into parallel and perpendicular components to the electric field as a function of the magnetization orientation within the (001)-plane. In each plot, $\theta$ denotes the angle of the magnetization measured from the electric field. The fits were done using the lowest order expansion of the second-order conductivity tensor given in Eq. \eqref{eq:xi_exp}. This tensor does not depend on $\mathbf{E}$ and thus the fitting is done together for all directions of $\textbf{E}$. The fitting is done for both the transverse and longitudinal components together. We plot the results as a ratio of the second-order current to the first order longitudinal current. This ratio depends linearly on the electric field, and the values are given  for an electric field that corresponds to a longitudinal current density of $10^{10}$ Am$^{-2}$.
}
\label{fig2}
\end{figure}

\begin{figure}[h]
\hspace*{-0cm}\includegraphics[width=.9\textwidth]{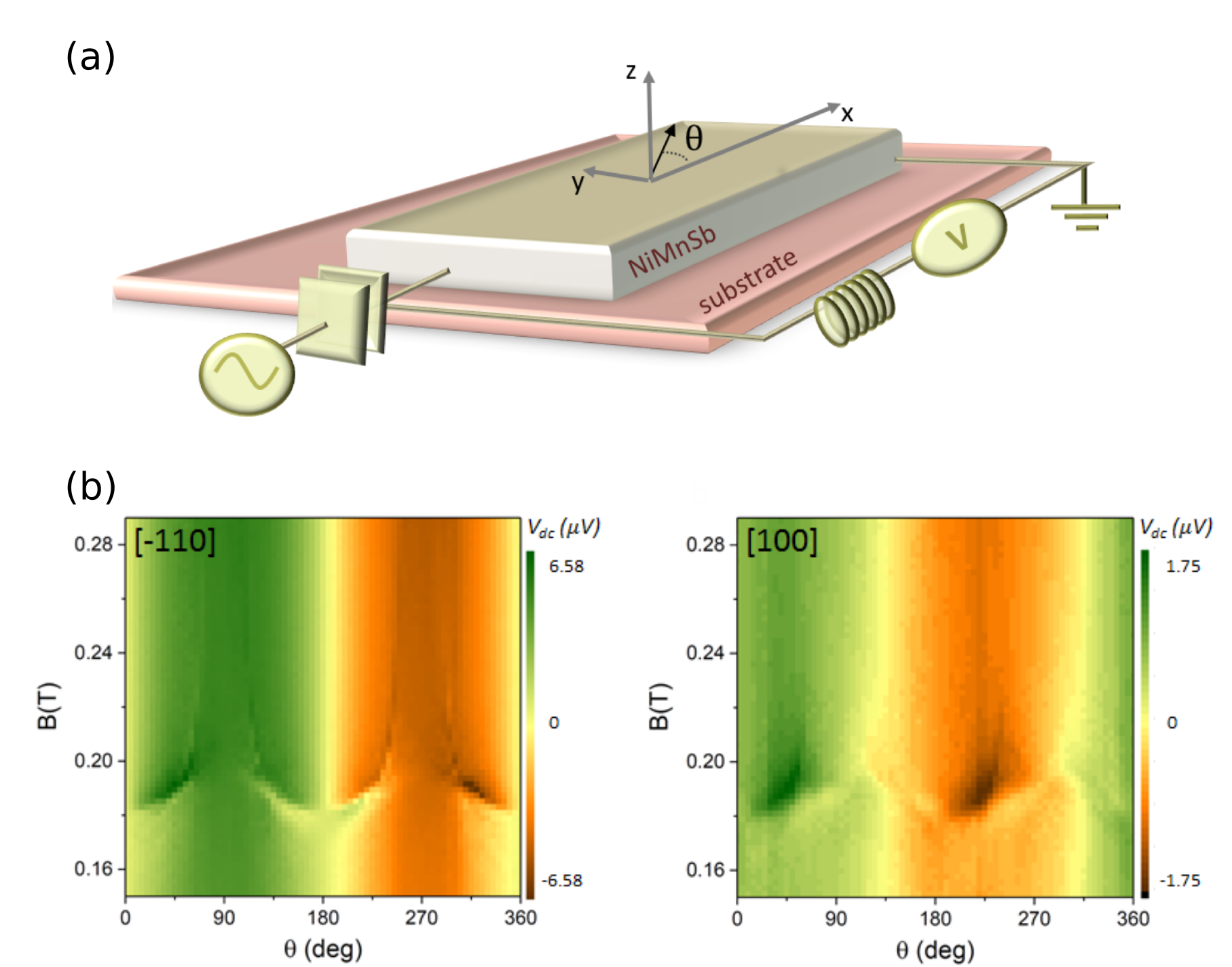}
\caption
{{\bf Experimental set-up and on and off-resonance measurements of the mixing dc signal.} {\bf (a)} Design of the experimental layout. \textbf{(b)} Longitudinal dc voltage measured as a function of the external field \textbf{B} and its direction $\theta$ with respect to the bar for a bar patterned along the [$\bar{1}$10]  and one patterned along the [100] direction. A current density of 5.7x10$^{10}$ Am$^{-2}$ was used for the [$\bar{1}$10] bar while a current density of 1.3x10$^{10}$ Am$^{-2}$ was used for the [100] bar. 
}
\label{fig3}
\end{figure}

\begin{figure}[h]
\hspace*{-0cm}\includegraphics[width=.9\textwidth]{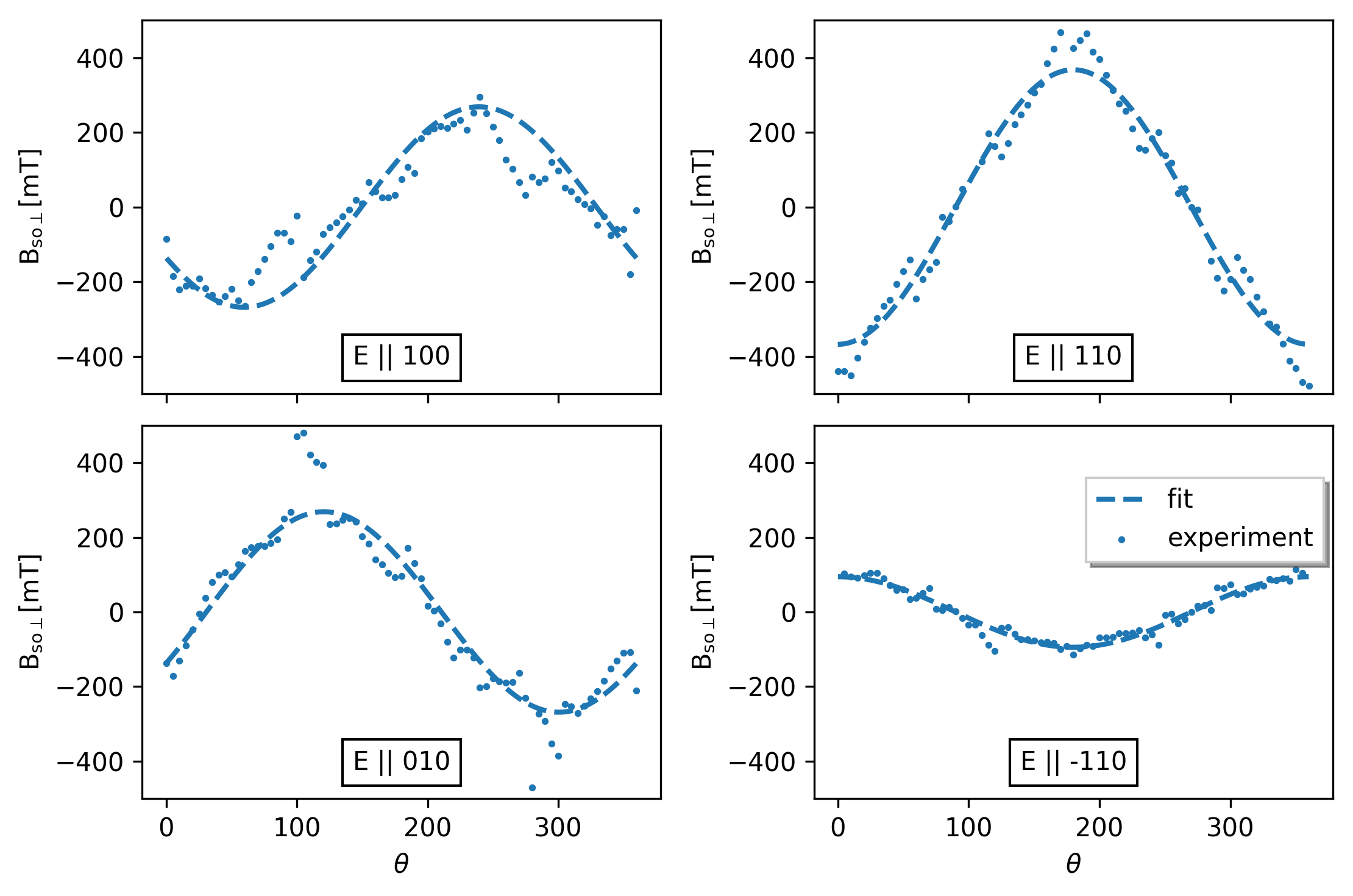}
\caption
{{\bf Experimental SOT measurements for different bar orientations.} The plots show the measured transverse components of the effective field $\mathbf{B}_\text{so}$ as a function of the magnetization orientation within the (001)-plane. In each plot $\theta$ denotes the angle of the magnetization measured from the electric field, as illustrated in Fig. \ref{fig3}(a). Dashed lines are the harmonic fits.
}
\label{fig4}
\end{figure}

\begin{figure}[h]
\hspace*{-0cm}\includegraphics[width=.9\textwidth]{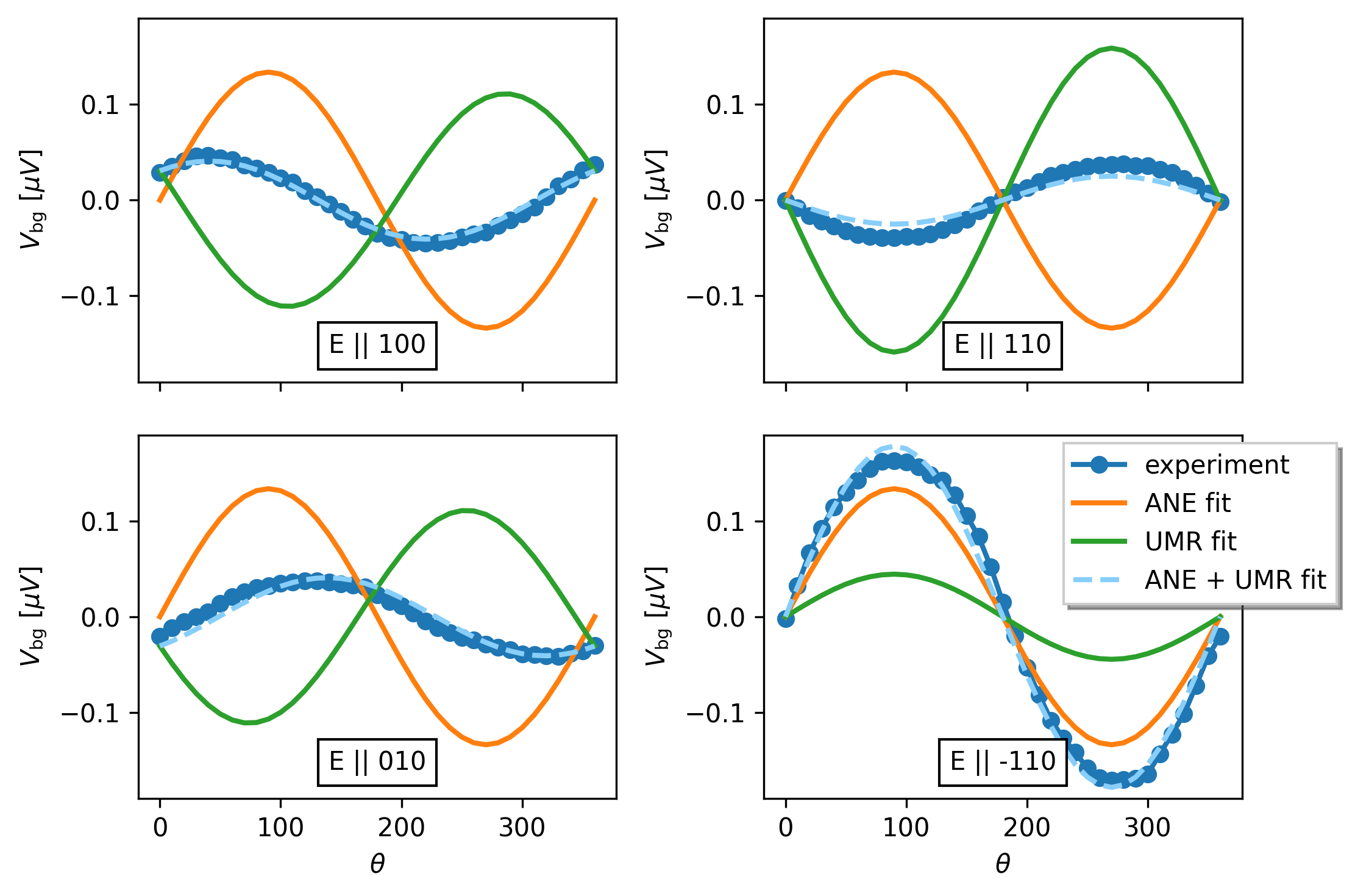}
\caption
{{\bf Extraction of the UMR and ANE contributions from the measured background voltage.} The plots show the measured background voltage as a function of the magnetization orientation within the (001)-plane. The UMR and ANE fits are done assuming that the ANE is isotropic and UMR is described by the lowest order second-order conductivity tensor  given in Eq. \eqref{eq:xi_exp} with the additional assumption that the relative amplitudes of the UMR for different bar directions are the same as the relative magnitudes of the SOT. This leaves three fitting parameters, which can be understood as the overall UMR magnitude, UMR phase shift for the [100] and [010] directions and the ANE magnitude.
}
\label{fig5}
\end{figure}


\clearpage
\appendix
\section{Second-order Boltzmann formula derivation}
\label{appendix:derivation}

Here we derive the second-order Boltzmann formula \eqref{Boltzmann}. The general form of the Boltzmann formula for a distribution function $g(t,\rb,\kb)$ under  the assumptions of a stationary  and spatially homogeneous $g$ and no magnetic field \cite{ziman1960} can be expressed as:

\begin{align}
 -\frac{e}{\hbar}\mathbf{E} \cdot \nabla_k g 
 = \left( \frac{dg}{dt} \right)_\text{col},
\end{align}
where $e$ is the (positive) elementary charge, $\mathbf{E}$ is the electric field and $\left( \frac{dg}{dt} \right)_\text{col}$ is the change of the distribution function due to scattering.

We will assume the constant relaxation-time approximation:
\begin{align}
    \left( \frac{dg}{dt} \right)_\text{col} = - \frac{g - g_0}{\tau},
\end{align}
where $\tau$ is the relaxation time and $g_0$ is the equilibrium distribution function, which for electrons is the Fermi-Dirac distribution function:
\begin{align}
        g_0(\rb,\kb) = f_{\text{FD}}(\calE (\kb)) = \frac{1}{e^{\frac{\calE(\kb)-\mu}{k_BT}}+1}.
\end{align}
Within the constant relaxation-time approximation the Boltzmann formula has the following form:
\begin{align}
 -\frac{e}{\hbar} \mathbf{E} \cdot \nabla_k g = - \frac{g - f_{\text{FD}}}{\tau}.\label{eq:Boltzmanm2}
\end{align}
To find a solution for $g$ up to second order in $E$ we expand $g$ in powers of $E$:
\begin{align}
 g = g_0 + g_1^i E_i + g_2^{ij}E_iE_j,
\end{align}
and insert it into the Boltzmanm formula \eqref{eq:Boltzmanm2}. Keeping only the terms up to $E^2$ we find:
\begin{align}
 -\frac{e}{\hbar} E_i \left( \frac{\partial g_0}{\partial k_i} + \frac{\partial}{\partial k_i} (g_1^j E_j) \right)a = - \frac{g_1^iE_i + g_2^{ij}E_iE_j}{\tau}.
\end{align}
Since this equation must hold for all $E$, the coefficients for the $E$ and $E^2$ terms on both sides of the equation must be the same. Therefore
\begin{align}
 g_1^i &= \frac{e\tau}{\hbar} \frac{\partial g_0}{\partial k_i}, \label{eq:g1} \\
 g_2^{ij} &= \frac{e\tau}{\hbar} \frac{\partial g_1^j}{\partial k_i} = \frac{e^2\tau^2}{\hbar^2} \frac{\partial^2 g_0}{\partial k_i \partial k_j}\label{eq:g2}.
\end{align}
Considering that the dependence of $g_0$ on $\kb$ is only through $\calE$ we find:
\begin{align}
 g_1^i &= \frac{e\tau}{\hbar}  \frac{\partial g_0}{\partial \calE } \frac{\partial \calE}{k_i},\\
 g_2^{ij} &= \frac{e^2\tau^2}{\hbar^2} \left( \frac{\partial^2 g_0}{\partial \calE^2} \frac{\partial \calE }{\partial k_i} \frac{\partial \calE }{\partial k_j} + \frac{\partial g_0}{\partial \calE } \frac{\partial^2 \cal E}{\partial k_i\partial k_j} \right).
\end{align}
Taking into account the relation
\begin{align}
 \mathbf{v}(\kb) = \frac{1}{\hbar}\frac{\partial \calE}{\partial \kb}
\end{align}
we have
\begin{align}
 g_1^i &= e\tau v^i \frac{\partial g_0}{\partial \calE } ,\\
 g_2^{ij} &= e^2\tau^2 \left( \frac{\partial^2 g_0}{\partial \calE^2} v^i v^j  + \frac{\partial g_0}{\partial \calE } \frac{1}{\hbar}\frac{\partial v^j}{\partial k_i} \right).
\end{align}
Electrical current is then given by:

\begin{align}
 \mathbf{J} = -e \int \frac{d \kb}{(2\pi)^3}\mathbf{v}(\kb)g.
\end{align}
We note that this integral is done over the first Brillouin zone or any other unit cell in the reciprocal space. The first order contribution to the current is given by
\begin{align}
 \mathbf{J}_1 = -e^2\tau E_i \int \frac{d \kb}{(2\pi)^3}\mathbf{v}v^i \frac{\partial g_0}{\partial \calE },
\end{align}
and the second order contribution:
\begin{align}
  \mathbf{J}_2 = -e^3\tau^2 E_iE_j \int \frac{d \kb}{(2\pi)^3}\mathbf{v} \left( v^iv^j \frac{\partial^2 g_0}{\partial \calE^2 } +\frac{1}{\hbar}\frac{\partial v^j}{\partial k_i} \frac{\partial g_0}{\partial \calE } \right).\label{eq:j2}
\end{align}
Note that in a multi-band system, these expression give a contribution from each individual band and the total current will be a sum over all bands.

The expression for the second order current can be further simplified. We first define a second order conductivity tensor $\xi_{ijl}$:

\begin{align}
J_2^i = \xi_{ijl}E_jE_l.\label{eq:sigma2_def}
\end{align}
From Eq. \eqref{eq:j2} we have

\begin{align}
 \xi_{ijl} = \xi_{ijl}^a + \xi_{ijl}^b,\label{eq:j2}
\end{align}
where 
\begin{align}
 \xi_{ijl}^a &= -e^3\tau^2  \int \frac{d \kb}{(2\pi)^3} v^i v^j v^l \frac{\partial^2 g_0}{\partial \calE^2 } \\
 \xi_{ijl}^b &=  -e^3\tau^2  \int \frac{d \kb}{(2\pi)^3} v^i\frac{1}{\hbar}\frac{\partial v^l}{\partial k_j} \frac{\partial g_0}{\partial \calE } .
\end{align}

\begin{figure}[h]
\hspace*{-0cm}\includegraphics[width=.6\textwidth]{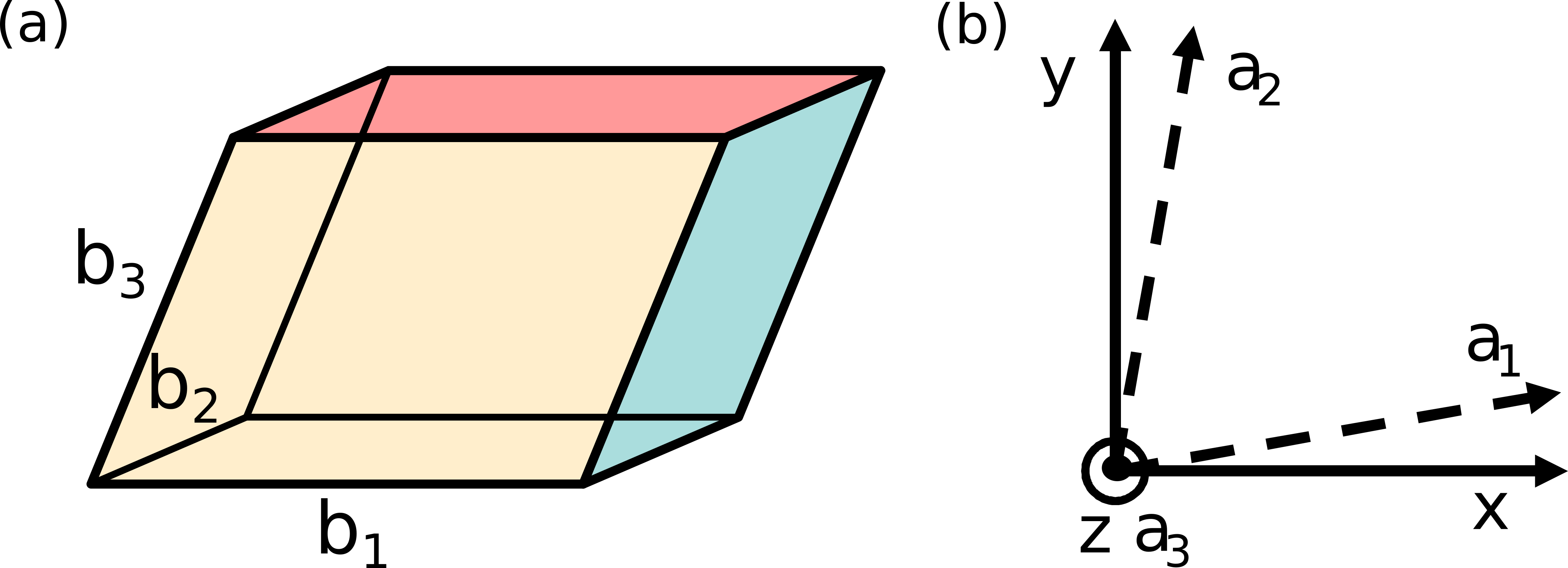}
\caption
{\textbf{(a)} Unit cell of the reciprocal space defined by the reciprocal lattice vectors $\mathbf{b}_1, \mathbf{b}_2, \mathbf{b}_3$. Image adapted from https://commons.wikimedia.org/w/index.php?curid=29922624 under CC BY-SA 3.0. \textbf{(b)} Coordinate system used for the symmetry analysis. The vectors $\textbf{x}$, $\textbf{y}$, $\textbf{z}$ define the cartesian coordinate system used for the symmetry analysis. $\textbf{a}_1$, $\textbf{a}_2$, $\textbf{a}_3$ are the lattice vectors of the NiMnSb in presence of shear strain. Without the strain, the lattice vectors would correspond to the conventional lattice of the cubic NiMnSb lattice and would be oriented along the cartesian coordinate system.}
\label{fig:Parallelepiped}
\end{figure}

Alternatively, using Eqs. \eqref{eq:g1} and \eqref{eq:g2}, $\xi$ can be written in the form
\begin{align}
  \xi_{ijl} = -\frac{e^3\tau^2}{\hbar} \int \frac{d \kb}{(2\pi)^3} v^i \frac{\partial }{\partial k_j} \left( v^l \frac{\partial g_0}{\partial \calE} \right),\label{eq:sigma2}
\end{align}
Using integration by parts
\begin{align}
  \xi_{ijl} = -\frac{e^3\tau^2}{\hbar} \int_\Gamma \frac{d k}{(2\pi)^3} v^i  v^l \nu^j \frac{\partial g_0}{\partial \calE} + \frac{e^3\tau^2}{\hbar} \int \frac{d \kb}{(2\pi)^3} \frac{\partial v^i}{\partial k_j} v^l \frac{\partial g_0}{\partial \calE},\label{eq:per_partez}
\end{align}
here $\Gamma$ signifies integral over the unit cell boundary and $\nu$ is the outward unit normal vector to the boundary. The first term in this relation in fact vanishes. To see that this is the case it is useful to use for the integration a unit cell of the reciprocal space spanned by the reciprocal lattice vectors $\mathbf{b}_1, \mathbf{b}_2, \mathbf{b}_3$ (see Fig. \eqref{fig:Parallelepiped}), instead of the first Brillouin zone. The first term in \eqref{eq:per_partez} is then given by sum over 6 surfaces. Since $g_0$ is a periodic function of $\kb$, also $\frac{\partial g_0}{\partial \calE}$ must be periodic. This means that at the opposite boundaries of the reciprocal unit cell $\frac{\partial g_0}{\partial \calE}$ is the same. However, since the outward unit normal vector $\nu$ is opposite for the opposite boundaries the whole term vanishes.

Combining Eqs. \eqref{eq:sigma2} and \eqref{eq:per_partez} we  have
\begin{align}
 \xi_{ijl} = \xi_{ijl}^a  + \xi_{ijl}^b = -\xi_{lji}^{b} \label{eq:sigma2_2}
\end{align}
To simplify this further we need to show that $\xi_{ijl}$ is symmetric under interchanging any two indices. This is clearly satisfied for $\xi_{ijl}^{a}$, however, it is less obvious for $\xi_{ijl}^{b}$. From Eq. \eqref{eq:sigma2_def}, we see that $\xi_{ijl} =  \xi_{ilj}$ must hold and thus also $\xi_{ijl}^{b} =  \xi_{ilj}^{2,b}$ should be satisfied. This can be explicitly verified from $\frac{1}{\hbar}\frac{\partial v^l}{\partial k_j}=\frac{\partial^2\calE}{\partial k_l \partial k_j}$, which is symmetric as long as $\calE$ is sufficiently smooth function.  To show that the tensors are also symmetric under interchanging the other indices, we consider:
\begin{align}
 \xi_{jil} - \xi_{ijl} =  -\xi_{lij}^{b} + \xi_{lji}^{b} = 0.
\end{align}
Then it must also hold that the $\xi_{ijl}$ and $\xi_{ijl}^b$ tensors are symmetric under interchanging indices $i$ and $l$:
\begin{align}
 \xi_{ijl} = \xi_{jil} = \xi_{jli} = \xi_{lji}.
\end{align}
 We can thus rewrite Eq. \eqref{eq:sigma2_2} as
\begin{align}
 \xi_{ijl} = \xi_{ijl}^{a}  + \xi_{ijl}^{b} = -\xi_{ijl}^{b}.
\end{align}
Therefore, $\xi_{ijl}^{b} = -\xi_{ijl}^{b}/2$  and  we finally find:
\begin{align}
 \xi_{ijl} = \frac{\xi_{ijl}^{a}}{2} = -\frac{e^3\tau^2}{2}  \int \frac{d \kb}{(2\pi)^3} v^i v^j v^l \frac{\partial^2 g_0}{\partial \calE^2 }.
\end{align}

\section{Symmetry of second order currents}
\label{appendix:symmetry}

Here we study the symmetry of the second-order currents. Similarly to other response phenomena, the second order currents will in general contain both time-reversal even ($\cal{T}$-even) and time-reversal odd ($\cal{T}$-odd) components. Here we consider only the time-reversal odd component since this component corresponds to the UMR. In the following $\xi_{ijk}$ will denote the $\cal{T}$-odd second-order conductivity tensor. Furthermore, we assume the $\xi_{ijk}$ tensor is symmetric under interchanging any two indices. As shown in Appendix \ref{appendix:derivation}, this holds for the Boltzmann formula that we use in our calculations. It is not clear whether this holds for the $\xi_{ijk}$ in general, thus the symmetry analysis here should be taken to refer specifically to the Boltzmann contribution. The method for the symmetry analysis is analogous to the one used in Ref. \onlinecite{Zelezny2017}. We have implemented the second order symmetry analysis in the open source code Symmetr. \cite{symcode} All the results shown here are given in a cartesian coordinate system described in Fig. \ref{fig:Parallelepiped}(b). Since in the experiment the magnetization always lies in the [001] plane, we consider here only magnetization in this plane. In Table \ref{table:symmetrytensors} we give the general shape of the $\xi_{ijk}$ for general direction of magnetization within this plane as well as for the [110] and [1-10] directions where the symmetry is higher. 

\begin{table}[h]
 \begin{tabular}{p{35mm}ccc}
   \hline \hline 
 \rule{0pt}{11pt}
       & $\xi_{xjl}$ & $\xi_{yjl}$ & $\xi_{zjl}$ \\
\hline
\noalign{\vskip 1mm}
 $\mathbf{M} \perp [001]$ & 
                $\left(\begin{matrix}\xi_{xxx} & \xi_{yxx} & 0\\\xi_{yxx} & \xi_{yyx} & 0\\0 & 0 & \xi_{zzx}\end{matrix}\right)$ &
                $\left(\begin{matrix}\xi_{yxx} & \xi_{yyx} & 0\\\xi_{yyx} & \xi_{yyy} & 0\\0 & 0 & \xi_{zzy}\end{matrix}\right)$ &
                $\left(\begin{matrix}0 & 0 & \xi_{zzx}\\0 & 0 & \xi_{zzy}\\\xi_{zzx} & \xi_{zzy} & 0\end{matrix}\right)$\\
 $\mathbf{M} || [110]$ or $[1-10]$ & 
                $\left(\begin{matrix}\xi_{xxx} & \xi_{yxx} & 0\\\xi_{yxx} & - \xi_{yxx} & 0\\0 & 0 & \xi_{zzx}\end{matrix}\right)$ &
                $\left(\begin{matrix}\xi_{yxx} & - \xi_{yxx} & 0\\- \xi_{yxx} & - \xi_{xxx} & 0\\0 & 0 & - \xi_{zzx}\end{matrix}\right)$ &
                $\left(\begin{matrix}0 & 0 & \xi_{zzx}\\0 & 0 & - \xi_{zzx}\\\xi_{zzx} & - \xi_{zzx} & 0\end{matrix}\right)$\\
                
 \noalign{\vskip 1mm}
  \hline \hline
\noalign{\vskip 1mm}

\label{table:symmetry_soc}
 \end{tabular}
\caption{Symmetry of second-order conductivity tensor in NiMnSb for different directions of the magnetization. Note that the $\xi_{ijk}$ coefficients between different directions of $\textbf{M}$ are not in general related.}
\label{table:symmetrytensors}
 \end{table}

To describe the dependence of the second-order currents on magnetization, it is useful to expand the $\xi_{ijk}$ tensor in powers of the magnetization. We consider only the lowest order term since it describes well both the calculations and the experiment:
\begin{align}
  \xi_{xjl}   &=  
    \left(\begin{matrix}M_{x} x_{1} - M_{y} x_{2} & M_{x} x_{3} - M_{y} x_{4} & M_{z} x_{5}\\M_{x} x_{3} - M_{y} x_{4} & M_{x} x_{4} - M_{y} x_{3} & 0\\M_{z} x_{5} & 0 & M_{x} x_{6} - M_{y} x_{7}\end{matrix}\right),\notag \\
 \xi_{yjl}   &= 
    \left(\begin{matrix}M_{x} x_{3} - M_{y} x_{4} & M_{x} x_{4} - M_{y} x_{3} & 0\\M_{x} x_{4} - M_{y} x_{3} & M_{x} x_{2} - M_{y} x_{1} & - M_{z} x_{5}\\0 & - M_{z} x_{5} & M_{x} x_{7} - M_{y} x_{6}\end{matrix}\right),\label{eq:xi_exp} \\
 \xi_{zjl}   &=
    \left(\begin{matrix}M_{z} x_{5} & 0 & M_{x} x_{6} - M_{y} x_{7}\\0 & - M_{z} x_{5} & M_{x} x_{7} - M_{y} x_{6}\\M_{x} x_{6} - M_{y} x_{7} & M_{x} x_{7} - M_{y} x_{6} & 0\end{matrix}\right).\notag
\end{align}
Here $x_i$ denotes free parameters of the expansion. 

\section{Calculation description}
\label{appendix:calculations}

The calculations utilize the FPLO density-functional theory (DFT) code\cite{FPLOpaper,FPLO} for description of the electronic structure. This DFT uses a local orbitals basis set for solving the Kohn-Sham equations. This makes it easy to transform the DFT Kohn-Sham Hamiltonian into a Wannier form, which is needed for the transport calculations. We use the full set of basis orbitals for this transformation, which makes the Wannier Hamiltonian very accurate. This is crucial for the second-order calculations since the second order contributions are very small and very sensitive to small symmetry violations that often exist in Wannier Hamiltonians generated by the more conventional approach based on maximally localized Wannier functions. The transport calculations utilize Wannier interpolation to evaluate the response formula on a tight grid in the reciprocal space. We have implemented the second-order Boltzmann calculation in the freely accessible Linres code.\cite{linear_response_code} We have also utilized this code for the calculations of the current-induced spin polarization. 

The spin-polarization calculations use the following Kubo formula for the esponse tensor (i.e., tensor such that  $\delta s_{\text{SO},i} = \chi_{ij}E_j$):\cite{freimuth2014}

 \begin{eqnarray}
	\chi_{ij} = 
	-\frac{e\hbar}{V\pi} \text{Re}\sum_{\textbf{k}, m, n}
	\frac{\braket{u_{n}(\textbf{k}) | \hat{S}_i | u_{m}(\textbf{k})}\Braket{ u_{m}(\textbf{k}) | \hat{v}_j | u_{n}(\textbf{k})} \Gamma^{2}}{(\left(E_{F}-E_{n}(\textbf{k}))^{2}+\Gamma^{2})(E_{F}-E_{m}(\textbf{k})\right)^{2}+\Gamma^{2})}\,. \nonumber
	\label{eq:Kubo_cisp}
\end{eqnarray}
Here, $\hat{\mathbf{S}}$ is the spin operator,   $u_n(\textbf{k})$ are the Bloch functions of a band $n$, \textbf{k} is the Bloch wave vector, $\varepsilon_{n}$(\textbf{k}) is the band energy, $E_F$ is the Fermi energy, $\hat{v}$ is the velocity operator and $\Gamma$ is a quasiparticle broadening parameter that describes the disorder strength. This formula becomes equivalent to the first-order constant relaxation time Boltzmann formula for small $\Gamma$ (with $\tau = \hbar/2\Gamma$), but differs for finite values of $\Gamma$. For the $\Gamma$ value that we consider the formula is close to the Boltzmann formula. We only consider the Boltzmann formula for the second order calculations since the derivation and evaluation of the second-order Kubo formula is much more complex. We note that the Kubo formula \eqref{eq:Kubo_cisp} describes only the $\cal{T}$-even part of the Kubo formula. We do not consider here the $\cal{T}$-odd component since experimentally it is seen that the in-plane SOT has a field-like character and the corresponding current-induced spin-polarization is thus $\cal{T}$-even. Our test calculations of the $\cal{T}$-odd component also suggest that it is much smaller than the $\cal{T}$-even component for realistic values of $\Gamma$.

The DFT calculations utilized 12x12x12 k-points and the GGA-PBE potential. The current-induced spin-polarization and the second-order conductivity response calculations use a 400x400x400 k-mesh, which we have confirmed to be sufficient for good convergence. To estimate the value of $\Gamma$ we calculate the first order conductivity using a conductivity formula analogous to Eq. \eqref{eq:Kubo_cisp} and choose the $\Gamma$ so that the conductivity matches the experimental conductivity. This corresponds to $\Gamma \approx$ 0.05 eV or alternatively to $\tau \approx$ 6.6 fs. In previous calculations of SOT in NiMnSb, a $\Gamma = 0.036$ eV was used,\cite{Ciccarelli2016} because the samples used in those experiments had somewhat larger conductivity. We note that our calculations of the current-induced spin-polarization were done for the $\Gamma = 0.036$ eV value and have been afterwords rescaled to $\Gamma = 0.05$ eV, assuming $1/\Gamma$ scaling. This is quite accurate since for theses value of $\Gamma$ the formula is very close to the Boltzmann formula and thus has $1/\Gamma$ scaling.

\section{Power dependence and reproducibility}
\label{appendix:power}
\begin{figure}[h]
\hspace*{-0cm}\includegraphics[width=.8\textwidth]{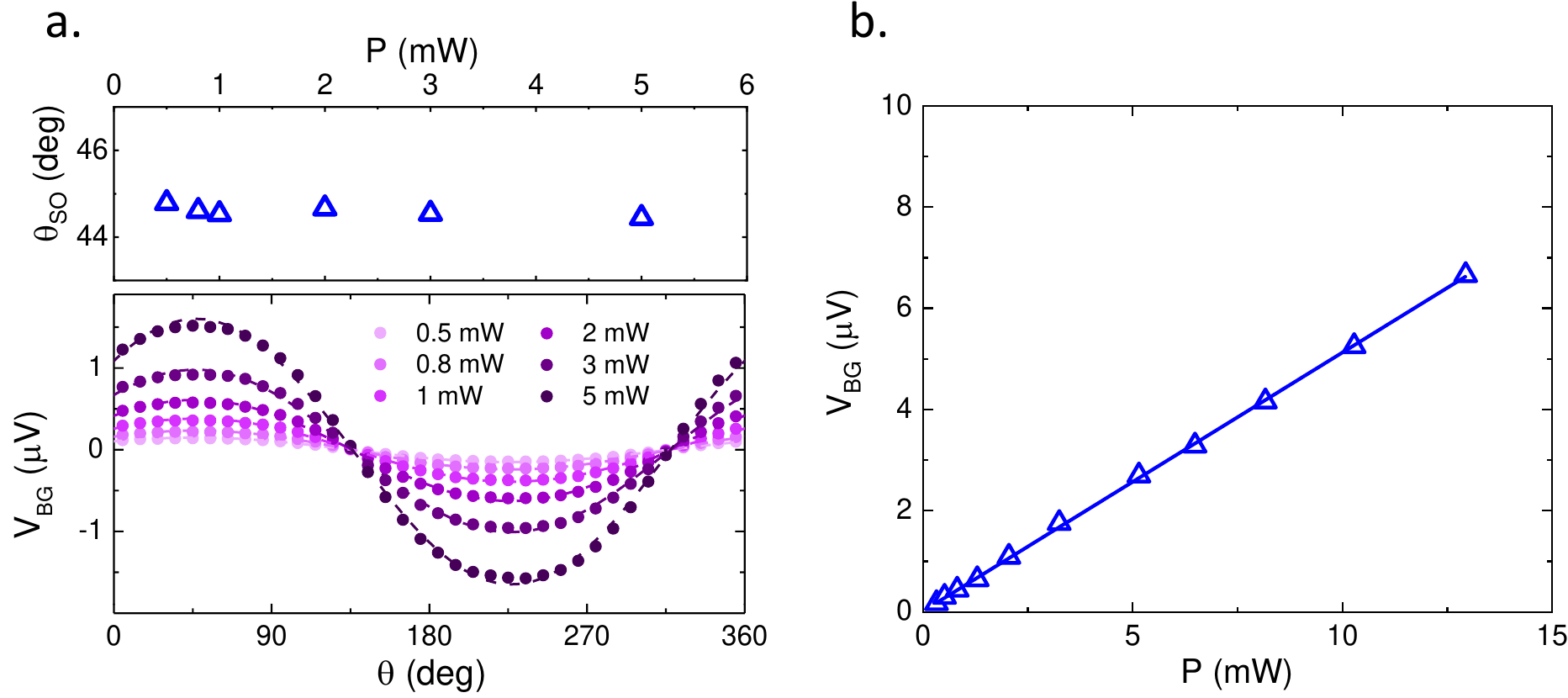}
\caption
{ \textbf{(a)} Top: Phase shift $\theta_{\text{SO}}$ of $V_\text{BG}$, fitted by $V_\text{SO}^0\sin(\theta+\theta_\text{SO})$, for a bar along [100] at different values of the microwave power. Bottom: angular dependence of $V_\text{BG}$ for different microwave powers passed in the bar, showing no phase shift. \textbf{(b)} Power dependence of $V_\text{BG}^0$ for the same bar. }
\label{fig:power_dependence}
\end{figure}

\begin{figure}[h]
\hspace*{-0cm}\includegraphics[width=\textwidth]{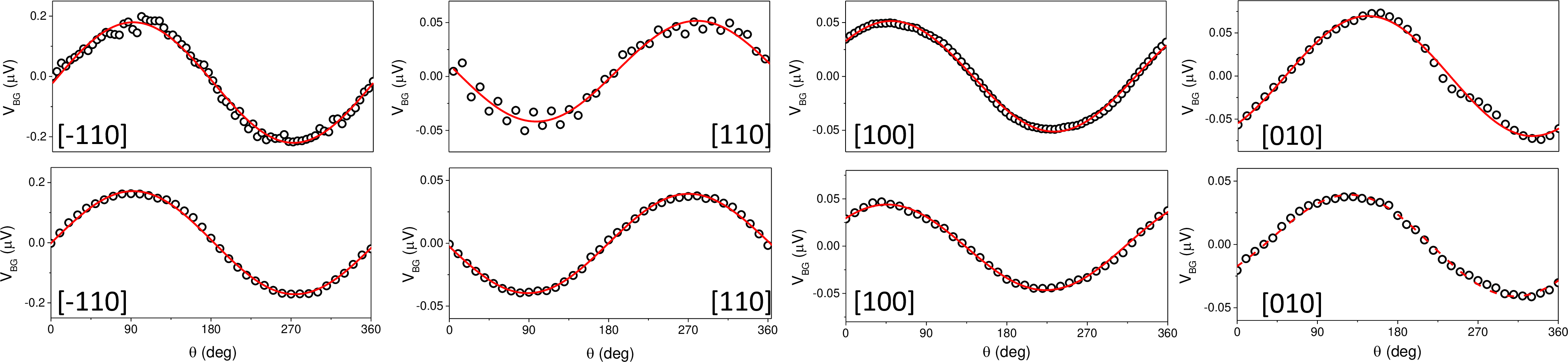}
\caption
{ Angular dependence of $V_\text{BG}$, normalized to a current density of $10^{10}$ Am$^{-2}$, measured for two different bars patterned on the same chip along each of the four crystal directions. Top: measurements for the first set of bars. Bottom: measurement for the second set of bars.}
\label{fig:reproducibility}
\end{figure}

In Fig. \ref{fig:power_dependence} we show that, as expected, the background voltage $V_\text{BG}$ scales linearly with power and that the angular dependence of $V_\text{BG}$ is independent of the power. In Fig. \ref{fig:reproducibility} we show that the results are reproducible between different bars patterned on the same chip.

\section{Modeling temperature gradient using Finite Element Method}
\label{appendix:ANE_modelling}

\begin{figure}[h]
\hspace*{-0cm}\includegraphics[width=.6\textwidth]{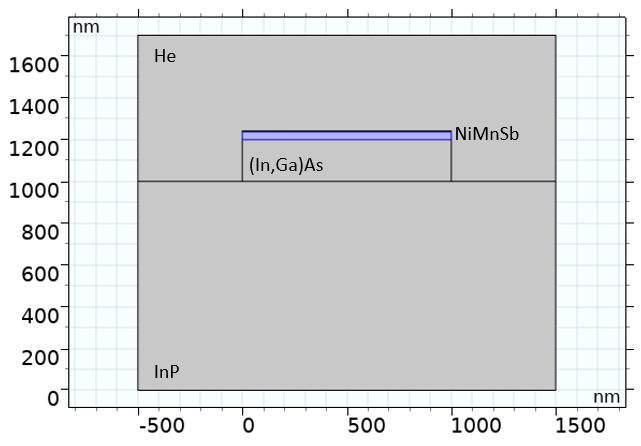}
\caption
{Simulated geometry - cross-section of the device, NiMnSn conducting the ac current is highlighted in blue, thin MgO capping layer (poorly visible in this plot) is included in the simulation.
}
\label{afig1}
\end{figure}

The system is excited by an ac current passing mainly through the NiMnSb layer which results in Joule heating. The heat scales with the square of the current density and dissipates into surrounding material giving rise to temperature gradients. The out-of-plane temperature gradient drives ANE which contributes to the electrical signal detected in the sample plane. We perform a simulation of heat transfer in the cross-section of our device utilizing the finite element method (FEM) as implemented in COMSOL Multiphysics\cite{comsol54}. The software solves the heat equation numerically using an automatically generated triangular mesh with density adjusted to the size of individual domains of the device. The geometry is shown in Figure~\ref{afig1} and consists of the NiMnSb wire (thickness = 37~nm, width = 1~$\mu$m) deposited on a (In,Ga)As  mesa (thickness = 200~nm, width = 1~$\mu$m). There is also an InP substrate (thickness = 1~$\mu$m, width= 2~$\mu$m), a thin MgO capping layer (thickness = 5~nm, width = 1~$\mu$m), and a He atmosphere included in the simulation. 

\begin{figure}[h]
\hspace*{-0cm}\includegraphics[width=.6\textwidth]{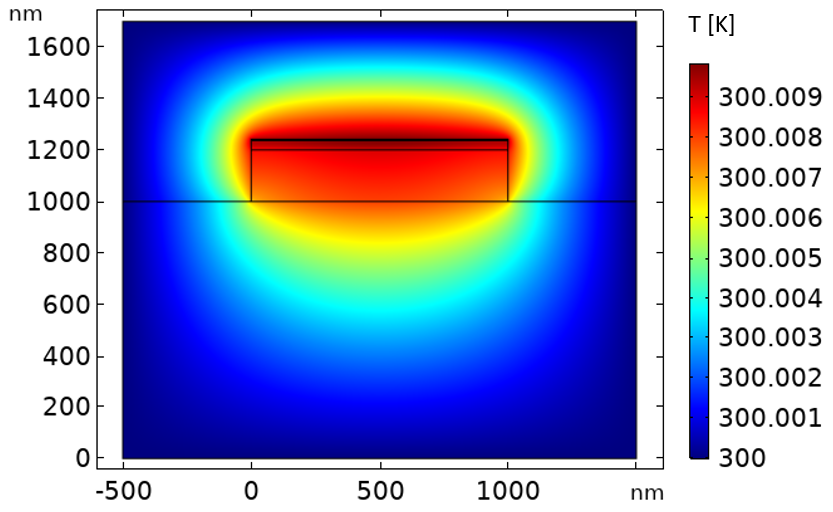}
\caption
{Temperature distribution in case the applied current density is 10$^{10}$ Am$^{-2}$ and all boundaries are set to T$_{ref}$ = 300~K – the most efficient cooling.
}
\label{afig2}
\end{figure}

\begin{figure}[h]
\hspace*{-0cm}\includegraphics[width=.6\textwidth]{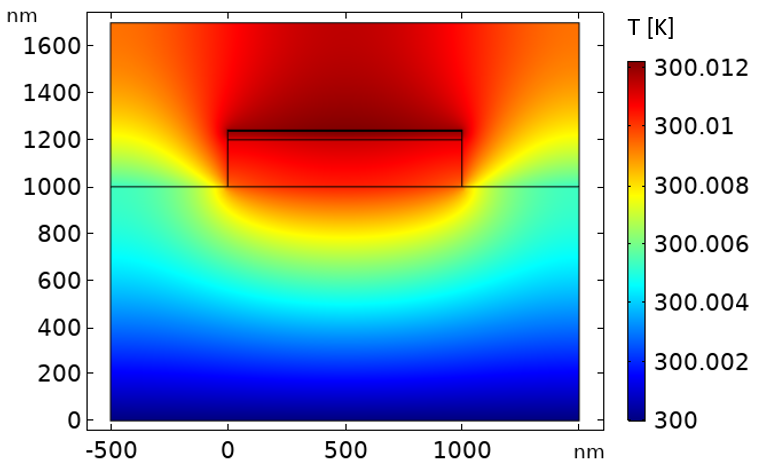}
\caption
{Temperature distribution in case the applied current density is 10$^{10}$ Am$^{-2}$ and  the top, left and right boundary of the simulated domain are insulating and the bottom boundary is set to T$_{ref}$ = 300~K - the least efficient cooling.
}
\label{afig3}
\end{figure}

The reference temperature, T$_{ref}$, is set to 300~K (before the Joule heating takes place). The current density applied to NiMnSb is 10$^{10}$ or 10$^{11}$~Am$^{-2}$. At the boundary of the simulated area the temperature is fixed to T$_{ref}$ or a thermal insulation is assumed (except the bottom boundary again fixed to T$_{ref}$). These two types of boundary condition correspond to very efficient cooling (transfer of heat to the surroundings) or to extremely poor cooling, respectively. The real system would fall between these two limiting cases. 
Figures~\ref{afig2} and~\ref{afig3} show the simulated steady-state temperature distribution for the case with boundaries fixed to T$_{ref}$ and with insulating boundaries, respectively. 

The former case (Fig.~\ref{afig2}) is simulated with applied current density of 10$^{10}$ and 10$^{11}$ Am$^{-2}$ but we show the temperature profile only for 10$^{10}$ Am$^{-2}$. The 10 times larger current density results visually in the same temperature profile but the maximum temperature increase, T-T$_{ref}$, is 100 times larger.

\begin{figure}[h]
\hspace*{-0cm}\includegraphics[width=.6\textwidth]{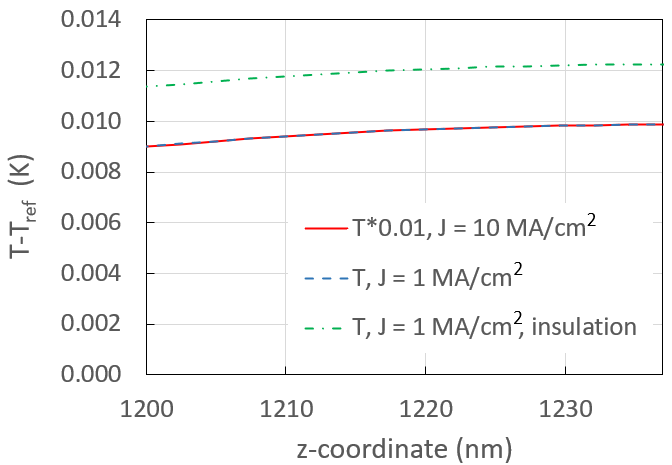}
\caption
{Temperature increase T-T$_{ref}$ along vertical cut-line withing the NiMnSb layer.
}
\label{afig4}
\end{figure}

The latter case (Fig.~\ref{afig3}) is simulated only with applied current density of 10$^{10}$ Am$^{-2}$ and all the heat is dissipated only via the bottom boundary – through the substrate. The out-of-plane temperature gradient is evaluated along a vertical cut-line (along the z-coordinate) running through the middle of the mesa.
Figure~\ref{afig4} shows the temperature increase, T-T$_{ref}$, along the cut-line within the NiMnSb layer (1200~nm to 1237~nm) for the three cases simulated. 

As expected, the case with insulating boundaries shows a larger increase in temperature but the dependence on the z-coordinate is the same. The case with 10 times larger applied current shows a 100 times larger increase of temperature which is due to the Joule heating scaling with the square of the applied current density.

\begin{figure}[h]
\hspace*{-0cm}\includegraphics[width=.6\textwidth]{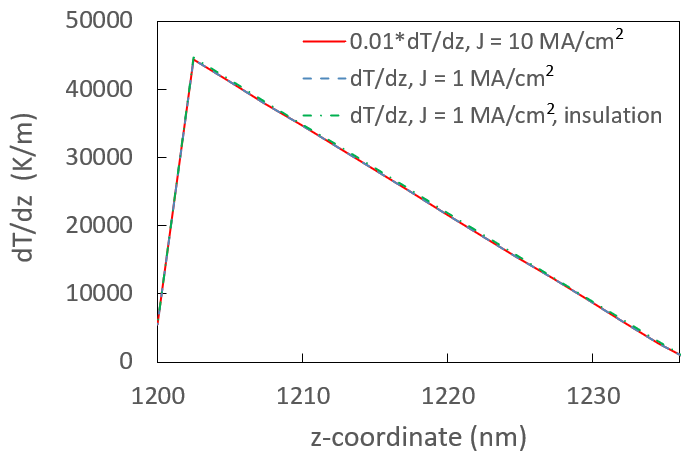}
\caption
{Out-of-plane temperature gradient along vertical cut-line within the NiMnSb layer.
}
\label{afig5}
\end{figure}

Finally, Figure~\ref{afig5} shows the out-of-plane temperature gradient generating ANE in the NiMnSb wire. It is evaluated simply as a numerical derivative of the temperature given in Figure~\ref{afig4} for the three cases. Note that the gradient decreases linearly in the NiMnSb layer towards the top surface. The efficiency of the cooling (insulating boundaries or fixed temperature) does not affect the gradient so the approximation of the cooling mechanism assumed in our model should not significantly compromise the validity of our numerical results. The main result of the FEM simulation relevant to the experimental device is that the average out-of-plane temperature gradient is of the order of 10$^4$~Km$^{-1}$ for current density of 10$^{10}$ Am$^{-2}$.

\begin{table}[h]
 \begin{tabular}{|c|c|c|c|c|}
   \hline
	        & $\sigma_E$ [Sm$^{-1}$] & $\sigma_T$ [Wm$^{-1}$K$^{-1}$] & $c_p$ [Jkg$^{-1}$K$^{-1}$] & $\rho$ [kgm$^{-3}$] \\
	 \hline
	 \hline
	 MgO    & 0        & 200\cite{slack1962thermal}   & 900\cite{MgO2020korth}   & 3600 \\
	 \hline
	 NiMnSb & 3.3e6    & 23\cite{gardelis2004synthesis}    & 420\cite{klicpera2020characterization}   & 7600 \\
	 \hline
	 InGaAs & 1.43e4   & 200\cite{carlson1965thermal}   & 300\cite{InGaAs2020ru}   & 5500 \\
	 \hline
	 InP    & 0        & 68\cite{InP2020ru}    & 310\cite{InP2020ru}   & 4800 \\
	 \hline
	 He     & 0        & 0.15\cite{He2020toolbox}  & 5200\cite{He2020toolbox}  & 145 \\
	 \hline
 \end{tabular}
\caption{Material parameters used as input of FEM model: electrical conductivity $\sigma_E$, thermal conductivity $\sigma_T$, heat capacity $c_p$, and mass density $\rho$.}
\label{atab1}
 \end{table}

Our FEM results depend on the material parameters used. We have measured the electrical conductivity of the individual layers at room temperature. The room temperature thermal conductivity and heat capacity parameters of our films are estimated based on literature as listed in Table~\ref{atab1}. In case of NiMnSb they are estimated based on related materials.\cite{gardelis2004synthesis,klicpera2020characterization} The thermal conductivity is underestimated, considering the metallic character of the film, so we report an upper estimate of the thermal gradient. The results are obtained for a steady state so there is no significant dependence on the heat capacity or mass density. (We estimate the mass density based on the molar masses and unit cell volume.)


\begin{thebibliography}{40}%
\makeatletter
\providecommand \@ifxundefined [1]{%
 \@ifx{#1\undefined}
}%
\providecommand \@ifnum [1]{%
 \ifnum #1\expandafter \@firstoftwo
 \else \expandafter \@secondoftwo
 \fi
}%
\providecommand \@ifx [1]{%
 \ifx #1\expandafter \@firstoftwo
 \else \expandafter \@secondoftwo
 \fi
}%
\providecommand \natexlab [1]{#1}%
\providecommand \enquote  [1]{``#1''}%
\providecommand \bibnamefont  [1]{#1}%
\providecommand \bibfnamefont [1]{#1}%
\providecommand \citenamefont [1]{#1}%
\providecommand \href@noop [0]{\@secondoftwo}%
\providecommand \href [0]{\begingroup \@sanitize@url \@href}%
\providecommand \@href[1]{\@@startlink{#1}\@@href}%
\providecommand \@@href[1]{\endgroup#1\@@endlink}%
\providecommand \@sanitize@url [0]{\catcode `\\12\catcode `\$12\catcode
  `\&12\catcode `\#12\catcode `\^12\catcode `\_12\catcode `\%12\relax}%
\providecommand \@@startlink[1]{}%
\providecommand \@@endlink[0]{}%
\providecommand \url  [0]{\begingroup\@sanitize@url \@url }%
\providecommand \@url [1]{\endgroup\@href {#1}{\urlprefix }}%
\providecommand \urlprefix  [0]{URL }%
\providecommand \Eprint [0]{\href }%
\providecommand \doibase [0]{http://dx.doi.org/}%
\providecommand \selectlanguage [0]{\@gobble}%
\providecommand \bibinfo  [0]{\@secondoftwo}%
\providecommand \bibfield  [0]{\@secondoftwo}%
\providecommand \translation [1]{[#1]}%
\providecommand \BibitemOpen [0]{}%
\providecommand \bibitemStop [0]{}%
\providecommand \bibitemNoStop [0]{.\EOS\space}%
\providecommand \EOS [0]{\spacefactor3000\relax}%
\providecommand \BibitemShut  [1]{\csname bibitem#1\endcsname}%
\let\auto@bib@innerbib\@empty
\bibitem [{\citenamefont {Mcguire}\ and\ \citenamefont
  {Potter}(1975)}]{McGuire1975}%
  \BibitemOpen
  \bibfield  {author} {\bibinfo {author} {\bibfnamefont {T.~R.}\ \bibnamefont
  {Mcguire}}\ and\ \bibinfo {author} {\bibfnamefont {R.~I.}\ \bibnamefont
  {Potter}},\ }\href {\doibase 10.1109/TMAG.1975.1058782} {\bibfield  {journal}
  {\bibinfo  {journal} {IEEE Transactions on Magnetics}\ }\textbf {\bibinfo
  {volume} {11}},\ \bibinfo {pages} {1018} (\bibinfo {year}
  {1975})}\BibitemShut {NoStop}%
\bibitem [{\citenamefont {Neumann}(1885)}]{Neumann1885}%
  \BibitemOpen
  \bibfield  {author} {\bibinfo {author} {\bibfnamefont {F.~E.}\ \bibnamefont
  {Neumann}},\ }in\ \href@noop {} {\emph {\bibinfo {booktitle} {Leipzig, B. G.
  Teubner-Verlag, edited by O. E. Meyer}}}\ (\bibinfo {year}
  {1885})\BibitemShut {NoStop}%
\bibitem [{\citenamefont {Bernevig}\ and\ \citenamefont
  {Zhang}(2005)}]{Bernevig2005a}%
  \BibitemOpen
  \bibfield  {author} {\bibinfo {author} {\bibfnamefont {B.~A.}\ \bibnamefont
  {Bernevig}}\ and\ \bibinfo {author} {\bibfnamefont {S.~C.}\ \bibnamefont
  {Zhang}},\ }\href {\doibase 10.1103/PhysRevB.72.115204} {\bibfield  {journal}
  {\bibinfo  {journal} {Physical Review B}\ }\textbf {\bibinfo {volume} {72}},\
  \bibinfo {pages} {115204} (\bibinfo {year} {2005})}\BibitemShut {NoStop}%
\bibitem [{\citenamefont {Chernyshov}\ \emph {et~al.}(2009)\citenamefont
  {Chernyshov}, \citenamefont {Overby}, \citenamefont {Liu}, \citenamefont
  {Furdyna}, \citenamefont {Lyanda-Geller},\ and\ \citenamefont
  {Rokhinson}}]{Chernyshov2009}%
  \BibitemOpen
  \bibfield  {author} {\bibinfo {author} {\bibfnamefont {A.}~\bibnamefont
  {Chernyshov}}, \bibinfo {author} {\bibfnamefont {M.}~\bibnamefont {Overby}},
  \bibinfo {author} {\bibfnamefont {X.}~\bibnamefont {Liu}}, \bibinfo {author}
  {\bibfnamefont {J.~K.}\ \bibnamefont {Furdyna}}, \bibinfo {author}
  {\bibfnamefont {Y.}~\bibnamefont {Lyanda-Geller}}, \ and\ \bibinfo {author}
  {\bibfnamefont {L.~P.}\ \bibnamefont {Rokhinson}},\ }\href {\doibase
  10.1038/nphys1362} {\bibfield  {journal} {\bibinfo  {journal} {Nature
  Physics}\ }\textbf {\bibinfo {volume} {5}},\ \bibinfo {pages} {656} (\bibinfo
  {year} {2009})}\BibitemShut {NoStop}%
\bibitem [{\citenamefont {Fang}\ \emph {et~al.}(2011)\citenamefont {Fang},
  \citenamefont {Kurebayashi}, \citenamefont {Wunderlich}, \citenamefont
  {V{\'{y}}born{\'{y}}}, \citenamefont {Z{\^{a}}rbo}, \citenamefont {Campion},
  \citenamefont {Casiraghi}, \citenamefont {Gallagher}, \citenamefont
  {Jungwirth},\ and\ \citenamefont {Ferguson}}]{Fang2011}%
  \BibitemOpen
  \bibfield  {author} {\bibinfo {author} {\bibfnamefont {D.}~\bibnamefont
  {Fang}}, \bibinfo {author} {\bibfnamefont {H.}~\bibnamefont {Kurebayashi}},
  \bibinfo {author} {\bibfnamefont {J.}~\bibnamefont {Wunderlich}}, \bibinfo
  {author} {\bibfnamefont {K.}~\bibnamefont {V{\'{y}}born{\'{y}}}}, \bibinfo
  {author} {\bibfnamefont {L.~P.}\ \bibnamefont {Z{\^{a}}rbo}}, \bibinfo
  {author} {\bibfnamefont {R.~P.}\ \bibnamefont {Campion}}, \bibinfo {author}
  {\bibfnamefont {A.}~\bibnamefont {Casiraghi}}, \bibinfo {author}
  {\bibfnamefont {B.~L.}\ \bibnamefont {Gallagher}}, \bibinfo {author}
  {\bibfnamefont {T.}~\bibnamefont {Jungwirth}}, \ and\ \bibinfo {author}
  {\bibfnamefont {A.~J.}\ \bibnamefont {Ferguson}},\ }\href {\doibase
  10.1038/nnano.2011.68} {\bibfield  {journal} {\bibinfo  {journal} {Nature
  Nanotechnology}\ }\textbf {\bibinfo {volume} {6}},\ \bibinfo {pages} {413}
  (\bibinfo {year} {2011})}\BibitemShut {NoStop}%
\bibitem [{\citenamefont {Kurebayashi}\ \emph {et~al.}(2014)\citenamefont
  {Kurebayashi}, \citenamefont {Sinova}, \citenamefont {Fang}, \citenamefont
  {Irvine}, \citenamefont {Skinner}, \citenamefont {Wunderlich}, \citenamefont
  {Nov{\'{a}}k}, \citenamefont {Campion}, \citenamefont {Gallagher},
  \citenamefont {Vehstedt}, \citenamefont {Z{\^{a}}rbo}, \citenamefont
  {V{\'{y}}born{\'{y}}}, \citenamefont {Ferguson},\ and\ \citenamefont
  {Jungwirth}}]{Kurebayashi2014}%
  \BibitemOpen
  \bibfield  {author} {\bibinfo {author} {\bibfnamefont {H.}~\bibnamefont
  {Kurebayashi}}, \bibinfo {author} {\bibfnamefont {J.}~\bibnamefont {Sinova}},
  \bibinfo {author} {\bibfnamefont {D.}~\bibnamefont {Fang}}, \bibinfo {author}
  {\bibfnamefont {A.~C.}\ \bibnamefont {Irvine}}, \bibinfo {author}
  {\bibfnamefont {T.~D.}\ \bibnamefont {Skinner}}, \bibinfo {author}
  {\bibfnamefont {J.}~\bibnamefont {Wunderlich}}, \bibinfo {author}
  {\bibfnamefont {V.}~\bibnamefont {Nov{\'{a}}k}}, \bibinfo {author}
  {\bibfnamefont {R.~P.}\ \bibnamefont {Campion}}, \bibinfo {author}
  {\bibfnamefont {B.~L.}\ \bibnamefont {Gallagher}}, \bibinfo {author}
  {\bibfnamefont {E.~K.}\ \bibnamefont {Vehstedt}}, \bibinfo {author}
  {\bibfnamefont {L.~P.}\ \bibnamefont {Z{\^{a}}rbo}}, \bibinfo {author}
  {\bibfnamefont {K.}~\bibnamefont {V{\'{y}}born{\'{y}}}}, \bibinfo {author}
  {\bibfnamefont {A.~J.}\ \bibnamefont {Ferguson}}, \ and\ \bibinfo {author}
  {\bibfnamefont {T.}~\bibnamefont {Jungwirth}},\ }\href {\doibase
  10.1038/nnano.2014.15} {\bibfield  {journal} {\bibinfo  {journal} {Nature
  Nanotechnology}\ }\textbf {\bibinfo {volume} {9}},\ \bibinfo {pages} {211}
  (\bibinfo {year} {2014})},\ \Eprint {http://arxiv.org/abs/1306.1893}
  {arXiv:1306.1893} \BibitemShut {NoStop}%
\bibitem [{\citenamefont {Ciccarelli}\ \emph {et~al.}(2015)\citenamefont
  {Ciccarelli}, \citenamefont {Hals}, \citenamefont {Irvine}, \citenamefont
  {Novak}, \citenamefont {Tserkovnyak}, \citenamefont {Kurebayashi},
  \citenamefont {Brataas},\ and\ \citenamefont {Ferguson}}]{Ciccarelli2014}%
  \BibitemOpen
  \bibfield  {author} {\bibinfo {author} {\bibfnamefont {C.}~\bibnamefont
  {Ciccarelli}}, \bibinfo {author} {\bibfnamefont {K.~M.}\ \bibnamefont
  {Hals}}, \bibinfo {author} {\bibfnamefont {A.}~\bibnamefont {Irvine}},
  \bibinfo {author} {\bibfnamefont {V.}~\bibnamefont {Novak}}, \bibinfo
  {author} {\bibfnamefont {Y.}~\bibnamefont {Tserkovnyak}}, \bibinfo {author}
  {\bibfnamefont {H.}~\bibnamefont {Kurebayashi}}, \bibinfo {author}
  {\bibfnamefont {A.}~\bibnamefont {Brataas}}, \ and\ \bibinfo {author}
  {\bibfnamefont {A.}~\bibnamefont {Ferguson}},\ }\href {\doibase
  10.1038/nnano.2014.252} {\bibfield  {journal} {\bibinfo  {journal} {Nature
  Nanotechnology}\ }\textbf {\bibinfo {volume} {10}},\ \bibinfo {pages} {50}
  (\bibinfo {year} {2015})}\BibitemShut {NoStop}%
\bibitem [{\citenamefont {Ciccarelli}\ \emph {et~al.}(2016)\citenamefont
  {Ciccarelli}, \citenamefont {Anderson}, \citenamefont {Tshitoyan},
  \citenamefont {Ferguson}, \citenamefont {Gerhard}, \citenamefont {Gould},
  \citenamefont {Molenkamp}, \citenamefont {Gayles}, \citenamefont
  {{\v{Z}}elezn{\'{y}}}, \citenamefont {{\v{S}}mejkal}, \citenamefont {Yuan},
  \citenamefont {Sinova}, \citenamefont {Freimuth},\ and\ \citenamefont
  {Jungwirth}}]{Ciccarelli2016}%
  \BibitemOpen
  \bibfield  {author} {\bibinfo {author} {\bibfnamefont {C.}~\bibnamefont
  {Ciccarelli}}, \bibinfo {author} {\bibfnamefont {L.}~\bibnamefont
  {Anderson}}, \bibinfo {author} {\bibfnamefont {V.}~\bibnamefont {Tshitoyan}},
  \bibinfo {author} {\bibfnamefont {A.~J.}\ \bibnamefont {Ferguson}}, \bibinfo
  {author} {\bibfnamefont {F.}~\bibnamefont {Gerhard}}, \bibinfo {author}
  {\bibfnamefont {C.}~\bibnamefont {Gould}}, \bibinfo {author} {\bibfnamefont
  {L.~W.}\ \bibnamefont {Molenkamp}}, \bibinfo {author} {\bibfnamefont
  {J.}~\bibnamefont {Gayles}}, \bibinfo {author} {\bibfnamefont
  {J.}~\bibnamefont {{\v{Z}}elezn{\'{y}}}}, \bibinfo {author} {\bibfnamefont
  {L.}~\bibnamefont {{\v{S}}mejkal}}, \bibinfo {author} {\bibfnamefont
  {Z.}~\bibnamefont {Yuan}}, \bibinfo {author} {\bibfnamefont {J.}~\bibnamefont
  {Sinova}}, \bibinfo {author} {\bibfnamefont {F.}~\bibnamefont {Freimuth}}, \
  and\ \bibinfo {author} {\bibfnamefont {T.}~\bibnamefont {Jungwirth}},\ }\href
  {\doibase 10.1038/nphys3772} {\bibfield  {journal} {\bibinfo  {journal}
  {Nature Physics}\ }\textbf {\bibinfo {volume} {12}},\ \bibinfo {pages} {855}
  (\bibinfo {year} {2016})}\BibitemShut {NoStop}%
\bibitem [{\citenamefont {Manchon}\ \emph {et~al.}(2019)\citenamefont
  {Manchon}, \citenamefont {{\v{Z}}elezn{\'{y}}}, \citenamefont {Miron},
  \citenamefont {Jungwirth}, \citenamefont {Sinova}, \citenamefont {Thiaville},
  \citenamefont {Garello},\ and\ \citenamefont {Gambardella}}]{Manchon2019}%
  \BibitemOpen
  \bibfield  {author} {\bibinfo {author} {\bibfnamefont {A.}~\bibnamefont
  {Manchon}}, \bibinfo {author} {\bibfnamefont {J.}~\bibnamefont
  {{\v{Z}}elezn{\'{y}}}}, \bibinfo {author} {\bibfnamefont {I.~M.}\
  \bibnamefont {Miron}}, \bibinfo {author} {\bibfnamefont {T.}~\bibnamefont
  {Jungwirth}}, \bibinfo {author} {\bibfnamefont {J.}~\bibnamefont {Sinova}},
  \bibinfo {author} {\bibfnamefont {A.}~\bibnamefont {Thiaville}}, \bibinfo
  {author} {\bibfnamefont {K.}~\bibnamefont {Garello}}, \ and\ \bibinfo
  {author} {\bibfnamefont {P.}~\bibnamefont {Gambardella}},\ }\href {\doibase
  10.1103/RevModPhys.91.035004} {\bibfield  {journal} {\bibinfo  {journal}
  {Reviews of Modern Physics}\ }\textbf {\bibinfo {volume} {91}},\ \bibinfo
  {pages} {035004} (\bibinfo {year} {2019})}\BibitemShut {NoStop}%
\bibitem [{\citenamefont {Silov}\ \emph {et~al.}(2004)\citenamefont {Silov},
  \citenamefont {Blajnov}, \citenamefont {Wolter}, \citenamefont {Hey},
  \citenamefont {Ploog},\ and\ \citenamefont {Averkiev}}]{Silov2004}%
  \BibitemOpen
  \bibfield  {author} {\bibinfo {author} {\bibfnamefont {A.~Y.}\ \bibnamefont
  {Silov}}, \bibinfo {author} {\bibfnamefont {P.~A.}\ \bibnamefont {Blajnov}},
  \bibinfo {author} {\bibfnamefont {J.~H.}\ \bibnamefont {Wolter}}, \bibinfo
  {author} {\bibfnamefont {R.}~\bibnamefont {Hey}}, \bibinfo {author}
  {\bibfnamefont {K.~H.}\ \bibnamefont {Ploog}}, \ and\ \bibinfo {author}
  {\bibfnamefont {N.~S.}\ \bibnamefont {Averkiev}},\ }\href {\doibase
  10.1063/1.1833565} {\bibfield  {journal} {\bibinfo  {journal} {Applied
  Physics Letters}\ }\textbf {\bibinfo {volume} {85}},\ \bibinfo {pages} {5929}
  (\bibinfo {year} {2004})}\BibitemShut {NoStop}%
\bibitem [{\citenamefont {Kato}\ \emph {et~al.}(2004)\citenamefont {Kato},
  \citenamefont {Myers}, \citenamefont {Gossard},\ and\ \citenamefont
  {Awschalom}}]{Kato2004b}%
  \BibitemOpen
  \bibfield  {author} {\bibinfo {author} {\bibfnamefont {Y.~K.}\ \bibnamefont
  {Kato}}, \bibinfo {author} {\bibfnamefont {R.~C.}\ \bibnamefont {Myers}},
  \bibinfo {author} {\bibfnamefont {A.~C.}\ \bibnamefont {Gossard}}, \ and\
  \bibinfo {author} {\bibfnamefont {D.~D.}\ \bibnamefont {Awschalom}},\ }\href
  {\doibase 10.1103/PhysRevLett.93.176601} {\bibfield  {journal} {\bibinfo
  {journal} {Physical Review Letters}\ }\textbf {\bibinfo {volume} {93}},\
  \bibinfo {pages} {176601} (\bibinfo {year} {2004})}\BibitemShut {NoStop}%
\bibitem [{\citenamefont {Ganichev}\ \emph {et~al.}(2004)\citenamefont
  {Ganichev}, \citenamefont {Danilov}, \citenamefont {Schneider}, \citenamefont
  {Bel'kov}, \citenamefont {Golub}, \citenamefont {Wegscheider}, \citenamefont
  {Weiss},\ and\ \citenamefont {Prettl}}]{Ganichev2004a}%
  \BibitemOpen
  \bibfield  {author} {\bibinfo {author} {\bibfnamefont {S.~D.}\ \bibnamefont
  {Ganichev}}, \bibinfo {author} {\bibfnamefont {S.~N.}\ \bibnamefont
  {Danilov}}, \bibinfo {author} {\bibfnamefont {P.}~\bibnamefont {Schneider}},
  \bibinfo {author} {\bibfnamefont {V.~V.}\ \bibnamefont {Bel'kov}}, \bibinfo
  {author} {\bibfnamefont {L.~E.}\ \bibnamefont {Golub}}, \bibinfo {author}
  {\bibfnamefont {W.}~\bibnamefont {Wegscheider}}, \bibinfo {author}
  {\bibfnamefont {D.}~\bibnamefont {Weiss}}, \ and\ \bibinfo {author}
  {\bibfnamefont {W.}~\bibnamefont {Prettl}},\ }\href
  {http://arxiv.org/abs/cond-mat/0403641} {\  (\bibinfo {year} {2004})},\
  \Eprint {http://arxiv.org/abs/0403641} {arXiv:0403641} \BibitemShut {NoStop}%
\bibitem [{\citenamefont {Wunderlich}\ \emph {et~al.}(2005)\citenamefont
  {Wunderlich}, \citenamefont {Kaestner}, \citenamefont {Sinova},\ and\
  \citenamefont {Jungwirth}}]{Wunderlich2005}%
  \BibitemOpen
  \bibfield  {author} {\bibinfo {author} {\bibfnamefont {J.}~\bibnamefont
  {Wunderlich}}, \bibinfo {author} {\bibfnamefont {B.}~\bibnamefont
  {Kaestner}}, \bibinfo {author} {\bibfnamefont {J.}~\bibnamefont {Sinova}}, \
  and\ \bibinfo {author} {\bibfnamefont {T.}~\bibnamefont {Jungwirth}},\ }\href
  {\doibase 10.1103/PhysRevLett.94.047204} {\bibfield  {journal} {\bibinfo
  {journal} {Physical Review Letters}\ }\textbf {\bibinfo {volume} {94}},\
  \bibinfo {pages} {047204} (\bibinfo {year} {2005})},\ \Eprint
  {http://arxiv.org/abs/0410295} {arXiv:0410295 [cond-mat]} \BibitemShut
  {NoStop}%
\bibitem [{\citenamefont {Sinova}\ \emph {et~al.}(2015)\citenamefont {Sinova},
  \citenamefont {Valenzuela}, \citenamefont {Wunderlich}, \citenamefont
  {Back},\ and\ \citenamefont {Jungwirth}}]{Sinova2015}%
  \BibitemOpen
  \bibfield  {author} {\bibinfo {author} {\bibfnamefont {J.}~\bibnamefont
  {Sinova}}, \bibinfo {author} {\bibfnamefont {S.~O.}\ \bibnamefont
  {Valenzuela}}, \bibinfo {author} {\bibfnamefont {J.}~\bibnamefont
  {Wunderlich}}, \bibinfo {author} {\bibfnamefont {C.~H.}\ \bibnamefont
  {Back}}, \ and\ \bibinfo {author} {\bibfnamefont {T.}~\bibnamefont
  {Jungwirth}},\ }\href {\doibase 10.1103/RevModPhys.87.1213} {\bibfield
  {journal} {\bibinfo  {journal} {Reviews of Modern Physics}\ }\textbf
  {\bibinfo {volume} {87}},\ \bibinfo {pages} {1213} (\bibinfo {year}
  {2015})}\BibitemShut {NoStop}%
\bibitem [{\citenamefont {Garello}\ \emph {et~al.}(2013)\citenamefont
  {Garello}, \citenamefont {Miron}, \citenamefont {Avci}, \citenamefont
  {Freimuth}, \citenamefont {Mokrousov}, \citenamefont {Bl{\"{u}}gel},
  \citenamefont {Auffret}, \citenamefont {Boulle}, \citenamefont {Gaudin},\
  and\ \citenamefont {Gambardella}}]{Garello2013}%
  \BibitemOpen
  \bibfield  {author} {\bibinfo {author} {\bibfnamefont {K.}~\bibnamefont
  {Garello}}, \bibinfo {author} {\bibfnamefont {I.~M.}\ \bibnamefont {Miron}},
  \bibinfo {author} {\bibfnamefont {C.~O.}\ \bibnamefont {Avci}}, \bibinfo
  {author} {\bibfnamefont {F.}~\bibnamefont {Freimuth}}, \bibinfo {author}
  {\bibfnamefont {Y.}~\bibnamefont {Mokrousov}}, \bibinfo {author}
  {\bibfnamefont {S.}~\bibnamefont {Bl{\"{u}}gel}}, \bibinfo {author}
  {\bibfnamefont {S.}~\bibnamefont {Auffret}}, \bibinfo {author} {\bibfnamefont
  {O.}~\bibnamefont {Boulle}}, \bibinfo {author} {\bibfnamefont
  {G.}~\bibnamefont {Gaudin}}, \ and\ \bibinfo {author} {\bibfnamefont
  {P.}~\bibnamefont {Gambardella}},\ }\href {\doibase 10.1038/nnano.2013.145}
  {\bibfield  {journal} {\bibinfo  {journal} {Nature Nanotechnology}\ }\textbf
  {\bibinfo {volume} {8}},\ \bibinfo {pages} {587} (\bibinfo {year}
  {2013})}\BibitemShut {NoStop}%
\bibitem [{\citenamefont {Avci}\ \emph {et~al.}(2015)\citenamefont {Avci},
  \citenamefont {Garello}, \citenamefont {Ghosh}, \citenamefont {Gabureac},
  \citenamefont {Alvarado},\ and\ \citenamefont {Gambardella}}]{Avci2015}%
  \BibitemOpen
  \bibfield  {author} {\bibinfo {author} {\bibfnamefont {C.~O.}\ \bibnamefont
  {Avci}}, \bibinfo {author} {\bibfnamefont {K.}~\bibnamefont {Garello}},
  \bibinfo {author} {\bibfnamefont {A.}~\bibnamefont {Ghosh}}, \bibinfo
  {author} {\bibfnamefont {M.}~\bibnamefont {Gabureac}}, \bibinfo {author}
  {\bibfnamefont {S.~F.}\ \bibnamefont {Alvarado}}, \ and\ \bibinfo {author}
  {\bibfnamefont {P.}~\bibnamefont {Gambardella}},\ }\href {\doibase
  10.1038/nphys3356} {\bibfield  {journal} {\bibinfo  {journal} {Nature
  Physics}\ }\textbf {\bibinfo {volume} {11}},\ \bibinfo {pages} {570}
  (\bibinfo {year} {2015})}\BibitemShut {NoStop}%
\bibitem [{\citenamefont {Olejn{\'{i}}k}\ \emph {et~al.}(2015)\citenamefont
  {Olejn{\'{i}}k}, \citenamefont {Nov{\'{a}}k}, \citenamefont {Wunderlich},\
  and\ \citenamefont {Jungwirth}}]{Olejnik2015}%
  \BibitemOpen
  \bibfield  {author} {\bibinfo {author} {\bibfnamefont {K.}~\bibnamefont
  {Olejn{\'{i}}k}}, \bibinfo {author} {\bibfnamefont {V.}~\bibnamefont
  {Nov{\'{a}}k}}, \bibinfo {author} {\bibfnamefont {J.}~\bibnamefont
  {Wunderlich}}, \ and\ \bibinfo {author} {\bibfnamefont {T.}~\bibnamefont
  {Jungwirth}},\ }\href {\doibase 10.1103/PhysRevB.91.180402} {\bibfield
  {journal} {\bibinfo  {journal} {Physical Review B}\ }\textbf {\bibinfo
  {volume} {91}},\ \bibinfo {pages} {1} (\bibinfo {year} {2015})}\BibitemShut
  {NoStop}%
\bibitem [{\citenamefont {Baibich}\ \emph {et~al.}(1988)\citenamefont
  {Baibich}, \citenamefont {Broto}, \citenamefont {Fert}, \citenamefont {{Van
  Dau}}, \citenamefont {Petroff}, \citenamefont {Eitenne}, \citenamefont
  {Creuzet}, \citenamefont {Friederich},\ and\ \citenamefont
  {Chazelas}}]{Baibich1988}%
  \BibitemOpen
  \bibfield  {author} {\bibinfo {author} {\bibfnamefont {M.~N.}\ \bibnamefont
  {Baibich}}, \bibinfo {author} {\bibfnamefont {J.~M.}\ \bibnamefont {Broto}},
  \bibinfo {author} {\bibfnamefont {A.}~\bibnamefont {Fert}}, \bibinfo {author}
  {\bibfnamefont {F.~N.}\ \bibnamefont {{Van Dau}}}, \bibinfo {author}
  {\bibfnamefont {F.}~\bibnamefont {Petroff}}, \bibinfo {author} {\bibfnamefont
  {P.}~\bibnamefont {Eitenne}}, \bibinfo {author} {\bibfnamefont
  {G.}~\bibnamefont {Creuzet}}, \bibinfo {author} {\bibfnamefont
  {A.}~\bibnamefont {Friederich}}, \ and\ \bibinfo {author} {\bibfnamefont
  {J.}~\bibnamefont {Chazelas}},\ }\href {\doibase 10.1103/PhysRevLett.61.2472}
  {\bibfield  {journal} {\bibinfo  {journal} {Physical Review Letters}\
  }\textbf {\bibinfo {volume} {61}},\ \bibinfo {pages} {2472} (\bibinfo {year}
  {1988})}\BibitemShut {NoStop}%
\bibitem [{\citenamefont {Binasch}\ \emph {et~al.}(1989)\citenamefont
  {Binasch}, \citenamefont {Gr{\"{u}}nberg}, \citenamefont {Saurenbach},\ and\
  \citenamefont {Zinn}}]{Binasch1989}%
  \BibitemOpen
  \bibfield  {author} {\bibinfo {author} {\bibfnamefont {G.}~\bibnamefont
  {Binasch}}, \bibinfo {author} {\bibfnamefont {P.}~\bibnamefont
  {Gr{\"{u}}nberg}}, \bibinfo {author} {\bibfnamefont {F.}~\bibnamefont
  {Saurenbach}}, \ and\ \bibinfo {author} {\bibfnamefont {W.}~\bibnamefont
  {Zinn}},\ }\href {\doibase 10.1103/PhysRevB.39.4828} {\bibfield  {journal}
  {\bibinfo  {journal} {Physical Review B}\ }\textbf {\bibinfo {volume} {39}},\
  \bibinfo {pages} {4828} (\bibinfo {year} {1989})}\BibitemShut {NoStop}%
\bibitem [{\citenamefont {Avci}\ \emph {et~al.}(2018)\citenamefont {Avci},
  \citenamefont {Mendil}, \citenamefont {Beach},\ and\ \citenamefont
  {Gambardella}}]{Avci2018}%
  \BibitemOpen
  \bibfield  {author} {\bibinfo {author} {\bibfnamefont {C.~O.}\ \bibnamefont
  {Avci}}, \bibinfo {author} {\bibfnamefont {J.}~\bibnamefont {Mendil}},
  \bibinfo {author} {\bibfnamefont {G.~S.}\ \bibnamefont {Beach}}, \ and\
  \bibinfo {author} {\bibfnamefont {P.}~\bibnamefont {Gambardella}},\ }\href
  {\doibase 10.1103/PhysRevLett.121.087207} {\bibfield  {journal} {\bibinfo
  {journal} {Physical Review Letters}\ }\textbf {\bibinfo {volume} {121}},\
  \bibinfo {pages} {87207} (\bibinfo {year} {2018})},\ \Eprint
  {http://arxiv.org/abs/1806.05305} {arXiv:1806.05305} \BibitemShut {NoStop}%
\bibitem [{\citenamefont {Langenfeld}\ \emph {et~al.}(2016)\citenamefont
  {Langenfeld}, \citenamefont {Tshitoyan}, \citenamefont {Fang}, \citenamefont
  {Wells}, \citenamefont {Moore},\ and\ \citenamefont
  {Ferguson}}]{Langenfeld2016}%
  \BibitemOpen
  \bibfield  {author} {\bibinfo {author} {\bibfnamefont {S.}~\bibnamefont
  {Langenfeld}}, \bibinfo {author} {\bibfnamefont {V.}~\bibnamefont
  {Tshitoyan}}, \bibinfo {author} {\bibfnamefont {Z.}~\bibnamefont {Fang}},
  \bibinfo {author} {\bibfnamefont {A.}~\bibnamefont {Wells}}, \bibinfo
  {author} {\bibfnamefont {T.~A.}\ \bibnamefont {Moore}}, \ and\ \bibinfo
  {author} {\bibfnamefont {A.~J.}\ \bibnamefont {Ferguson}},\ }\href {\doibase
  10.1063/1.4948921} {\bibfield  {journal} {\bibinfo  {journal} {Applied
  Physics Letters}\ }\textbf {\bibinfo {volume} {108}},\ \bibinfo {pages}
  {192402} (\bibinfo {year} {2016})}\BibitemShut {NoStop}%
\bibitem [{\citenamefont {Zhang}\ \emph {et~al.}(2018)\citenamefont {Zhang},
  \citenamefont {{\v{Z}}elezn{\'{y}}}, \citenamefont {Sun}, \citenamefont
  {van~den Brink},\ and\ \citenamefont {Yan}}]{Zhang2018h}%
  \BibitemOpen
  \bibfield  {author} {\bibinfo {author} {\bibfnamefont {Y.}~\bibnamefont
  {Zhang}}, \bibinfo {author} {\bibfnamefont {J.}~\bibnamefont
  {{\v{Z}}elezn{\'{y}}}}, \bibinfo {author} {\bibfnamefont {Y.}~\bibnamefont
  {Sun}}, \bibinfo {author} {\bibfnamefont {J.}~\bibnamefont {van~den Brink}},
  \ and\ \bibinfo {author} {\bibfnamefont {B.}~\bibnamefont {Yan}},\ }\href
  {\doibase 10.1088/1367-2630/aad1eb} {\bibfield  {journal} {\bibinfo
  {journal} {New Journal of Physics}\ }\textbf {\bibinfo {volume} {20}},\
  \bibinfo {pages} {073028} (\bibinfo {year} {2018})}\BibitemShut {NoStop}%
\bibitem [{\citenamefont {Gerhard}\ \emph {et~al.}(2014)\citenamefont
  {Gerhard}, \citenamefont {Schumacher}, \citenamefont {Gould},\ and\
  \citenamefont {Molenkamp}}]{Gerhard2014a}%
  \BibitemOpen
  \bibfield  {author} {\bibinfo {author} {\bibfnamefont {F.}~\bibnamefont
  {Gerhard}}, \bibinfo {author} {\bibfnamefont {C.}~\bibnamefont {Schumacher}},
  \bibinfo {author} {\bibfnamefont {C.}~\bibnamefont {Gould}}, \ and\ \bibinfo
  {author} {\bibfnamefont {L.~W.}\ \bibnamefont {Molenkamp}},\ }\href {\doibase
  10.1063/1.4867298} {\bibfield  {journal} {\bibinfo  {journal} {Journal of
  Applied Physics}\ }\textbf {\bibinfo {volume} {115}},\ \bibinfo {pages}
  {094505} (\bibinfo {year} {2014})}\BibitemShut {NoStop}%
\bibitem [{\citenamefont {Sharma}\ \emph {et~al.}(2019)\citenamefont {Sharma},
  \citenamefont {Wen}, \citenamefont {Takanashi},\ and\ \citenamefont
  {Mizuguchi}}]{Sharma2019}%
  \BibitemOpen
  \bibfield  {author} {\bibinfo {author} {\bibfnamefont {H.}~\bibnamefont
  {Sharma}}, \bibinfo {author} {\bibfnamefont {Z.}~\bibnamefont {Wen}},
  \bibinfo {author} {\bibfnamefont {K.}~\bibnamefont {Takanashi}}, \ and\
  \bibinfo {author} {\bibfnamefont {M.}~\bibnamefont {Mizuguchi}},\ }\href
  {\doibase 10.7567/1347-4065/aafe68} {\bibfield  {journal} {\bibinfo
  {journal} {Japanese Journal of Applied Physics}\ }\textbf {\bibinfo {volume}
  {58}},\ \bibinfo {pages} {SBBI03} (\bibinfo {year} {2019})}\BibitemShut
  {NoStop}%
\bibitem [{\citenamefont {Ziman}(1960)}]{ziman1960}%
  \BibitemOpen
  \bibfield  {author} {\bibinfo {author} {\bibfnamefont {J.~M.}\ \bibnamefont
  {Ziman}},\ }\href@noop {} {\emph {\bibinfo {title} {Electrons and phonons:
  the theory of transport phenomena in solids}}}\ (\bibinfo  {publisher}
  {Clarendon Press},\ \bibinfo {year} {1960})\BibitemShut {NoStop}%
\bibitem [{\citenamefont {{\v{Z}}elezn{\'{y}}}\ \emph
  {et~al.}(2017)\citenamefont {{\v{Z}}elezn{\'{y}}}, \citenamefont {Gao},
  \citenamefont {Manchon}, \citenamefont {Freimuth}, \citenamefont {Mokrousov},
  \citenamefont {Zemen}, \citenamefont {Ma{\v{s}}ek}, \citenamefont {Sinova},\
  and\ \citenamefont {Jungwirth}}]{Zelezny2017}%
  \BibitemOpen
  \bibfield  {author} {\bibinfo {author} {\bibfnamefont {J.}~\bibnamefont
  {{\v{Z}}elezn{\'{y}}}}, \bibinfo {author} {\bibfnamefont {H.}~\bibnamefont
  {Gao}}, \bibinfo {author} {\bibfnamefont {A.}~\bibnamefont {Manchon}},
  \bibinfo {author} {\bibfnamefont {F.}~\bibnamefont {Freimuth}}, \bibinfo
  {author} {\bibfnamefont {Y.}~\bibnamefont {Mokrousov}}, \bibinfo {author}
  {\bibfnamefont {J.}~\bibnamefont {Zemen}}, \bibinfo {author} {\bibfnamefont
  {J.}~\bibnamefont {Ma{\v{s}}ek}}, \bibinfo {author} {\bibfnamefont
  {J.}~\bibnamefont {Sinova}}, \ and\ \bibinfo {author} {\bibfnamefont
  {T.}~\bibnamefont {Jungwirth}},\ }\href {\doibase 10.1103/PhysRevB.95.014403}
  {\bibfield  {journal} {\bibinfo  {journal} {Phys. Rev. B}\ }\textbf {\bibinfo
  {volume} {95}},\ \bibinfo {pages} {014403} (\bibinfo {year}
  {2017})}\BibitemShut {NoStop}%
\bibitem [{\citenamefont {{{\v Z}elezn{\'y}}}({\natexlab{a}})}]{symcode}%
  \BibitemOpen
  \bibfield  {author} {\bibinfo {author} {\bibfnamefont {J.}~\bibnamefont {{{\v
  Z}elezn{\'y}}}},\ }\href@noop {} {}\bibinfo {howpublished}
  {\url{https://bitbucket.org/zeleznyj/linear-response-symmetry}}
  ({\natexlab{a}})\BibitemShut {NoStop}%
\bibitem [{\citenamefont {Koepernik}\ and\ \citenamefont
  {Eschrig}(1999)}]{FPLOpaper}%
  \BibitemOpen
  \bibfield  {author} {\bibinfo {author} {\bibfnamefont {K.}~\bibnamefont
  {Koepernik}}\ and\ \bibinfo {author} {\bibfnamefont {H.}~\bibnamefont
  {Eschrig}},\ }\href {\doibase 10.1103/PhysRevB.59.1743} {\bibfield  {journal}
  {\bibinfo  {journal} {Phys. Rev. B}\ }\textbf {\bibinfo {volume} {59}},\
  \bibinfo {pages} {1743} (\bibinfo {year} {1999})}\BibitemShut {NoStop}%
\bibitem [{FPL()}]{FPLO}%
  \BibitemOpen
  \href@noop {} {}\bibinfo {howpublished}
  {\url{https://www.fplo.de/}}\BibitemShut {NoStop}%
\bibitem [{\citenamefont {{{\v
  Z}elezn{\'y}}}({\natexlab{b}})}]{linear_response_code}%
  \BibitemOpen
  \bibfield  {author} {\bibinfo {author} {\bibfnamefont {J.}~\bibnamefont {{{\v
  Z}elezn{\'y}}}},\ }\href@noop {} {}\bibinfo {howpublished}
  {\url{https://bitbucket.org/zeleznyj/wannier-linear-response}}
  ({\natexlab{b}})\BibitemShut {NoStop}%
\bibitem [{\citenamefont {Freimuth}\ \emph {et~al.}(2014)\citenamefont
  {Freimuth}, \citenamefont {Bl{\"{u}}gel},\ and\ \citenamefont
  {Mokrousov}}]{freimuth2014}%
  \BibitemOpen
  \bibfield  {author} {\bibinfo {author} {\bibfnamefont {F.}~\bibnamefont
  {Freimuth}}, \bibinfo {author} {\bibfnamefont {S.}~\bibnamefont
  {Bl{\"{u}}gel}}, \ and\ \bibinfo {author} {\bibfnamefont {Y.}~\bibnamefont
  {Mokrousov}},\ }\href {\doibase 10.1088/0953-8984/26/10/104202} {\bibfield
  {journal} {\bibinfo  {journal} {Journal of Physics Condensed Matter}\
  }\textbf {\bibinfo {volume} {26}},\ \bibinfo {pages} {104202} (\bibinfo
  {year} {2014})}\BibitemShut {NoStop}%
\bibitem [{com()}]{comsol54}%
  \BibitemOpen
  \href@noop {} {\enquote {\bibinfo {title} {{COMSOL Multiphysics v. 5.4}},}\
  }\bibinfo {howpublished} {\url{www.comsol.com}},\ \bibinfo {note} {{COMSOL
  AB, Stockholm, Sweden.}}\BibitemShut {Stop}%
\bibitem [{\citenamefont {Slack}(1962)}]{slack1962thermal}%
  \BibitemOpen
  \bibfield  {author} {\bibinfo {author} {\bibfnamefont {G.~A.}\ \bibnamefont
  {Slack}},\ }\href {\doibase 10.1103/PhysRev.126.427} {\bibfield  {journal}
  {\bibinfo  {journal} {Phys. Rev.}\ }\textbf {\bibinfo {volume} {126}},\
  \bibinfo {pages} {427} (\bibinfo {year} {1962})}\BibitemShut {NoStop}%
\bibitem [{MgO()}]{MgO2020korth}%
  \BibitemOpen
  \href@noop {} {\enquote {\bibinfo {title} {{Magnesium Oxide}},}\ }\bibinfo
  {howpublished} {\url{www.korth.de/index.php/162/items/22.html}}\BibitemShut
  {NoStop}%
\bibitem [{\citenamefont {Gardelis}\ \emph {et~al.}(2004)\citenamefont
  {Gardelis}, \citenamefont {Androulakis}, \citenamefont {Migiakis},
  \citenamefont {Giapintzakis}, \citenamefont {Clowes}, \citenamefont
  {Bugoslavsky}, \citenamefont {Branford}, \citenamefont {Miyoshi},\ and\
  \citenamefont {Cohen}}]{gardelis2004synthesis}%
  \BibitemOpen
  \bibfield  {author} {\bibinfo {author} {\bibfnamefont {S.}~\bibnamefont
  {Gardelis}}, \bibinfo {author} {\bibfnamefont {J.}~\bibnamefont
  {Androulakis}}, \bibinfo {author} {\bibfnamefont {P.}~\bibnamefont
  {Migiakis}}, \bibinfo {author} {\bibfnamefont {J.}~\bibnamefont
  {Giapintzakis}}, \bibinfo {author} {\bibfnamefont {S.}~\bibnamefont
  {Clowes}}, \bibinfo {author} {\bibfnamefont {Y.}~\bibnamefont {Bugoslavsky}},
  \bibinfo {author} {\bibfnamefont {W.}~\bibnamefont {Branford}}, \bibinfo
  {author} {\bibfnamefont {Y.}~\bibnamefont {Miyoshi}}, \ and\ \bibinfo
  {author} {\bibfnamefont {L.}~\bibnamefont {Cohen}},\ }\href@noop {}
  {\bibfield  {journal} {\bibinfo  {journal} {Journal of applied physics}\
  }\textbf {\bibinfo {volume} {95}},\ \bibinfo {pages} {8063} (\bibinfo {year}
  {2004})}\BibitemShut {NoStop}%
\bibitem [{\citenamefont {Klicpera}\ \emph {et~al.}(2020)\citenamefont
  {Klicpera}, \citenamefont {Kratochv{\'\i}lov{\'a}}, \citenamefont {Kovaliuk},
  \citenamefont {Valenta},\ and\ \citenamefont
  {Colman}}]{klicpera2020characterization}%
  \BibitemOpen
  \bibfield  {author} {\bibinfo {author} {\bibfnamefont {M.}~\bibnamefont
  {Klicpera}}, \bibinfo {author} {\bibfnamefont {M.}~\bibnamefont
  {Kratochv{\'\i}lov{\'a}}}, \bibinfo {author} {\bibfnamefont {T.}~\bibnamefont
  {Kovaliuk}}, \bibinfo {author} {\bibfnamefont {J.}~\bibnamefont {Valenta}}, \
  and\ \bibinfo {author} {\bibfnamefont {R.}~\bibnamefont {Colman}},\
  }\href@noop {} {\bibfield  {journal} {\bibinfo  {journal} {Journal of
  Magnetism and Magnetic Materials}\ }\textbf {\bibinfo {volume} {513}},\
  \bibinfo {pages} {167083} (\bibinfo {year} {2020})}\BibitemShut {NoStop}%
\bibitem [{\citenamefont {Carlson}\ \emph {et~al.}(1965)\citenamefont
  {Carlson}, \citenamefont {Slack},\ and\ \citenamefont
  {Silverman}}]{carlson1965thermal}%
  \BibitemOpen
  \bibfield  {author} {\bibinfo {author} {\bibfnamefont {R.}~\bibnamefont
  {Carlson}}, \bibinfo {author} {\bibfnamefont {G.}~\bibnamefont {Slack}}, \
  and\ \bibinfo {author} {\bibfnamefont {S.}~\bibnamefont {Silverman}},\
  }\href@noop {} {\bibfield  {journal} {\bibinfo  {journal} {Journal of Applied
  Physics}\ }\textbf {\bibinfo {volume} {36}},\ \bibinfo {pages} {505}
  (\bibinfo {year} {1965})}\BibitemShut {NoStop}%
\bibitem [{InG()}]{InGaAs2020ru}%
  \BibitemOpen
  \href@noop {} {\enquote {\bibinfo {title} {{InGaAs}},}\ }\bibinfo
  {howpublished}
  {\url{www.ioffe.ru/SVA/NSM/Semicond/GaInAs/thermal.html}}\BibitemShut
  {NoStop}%
\bibitem [{InP()}]{InP2020ru}%
  \BibitemOpen
  \href@noop {} {\enquote {\bibinfo {title} {{InGaAs}},}\ }\bibinfo
  {howpublished}
  {\url{www.ioffe.ru/SVA/NSM/Semicond/InP/thermal.html}}\BibitemShut {NoStop}%
\bibitem [{He2()}]{He2020toolbox}%
  \BibitemOpen
  \href@noop {} {\enquote {\bibinfo {title} {{Helium}},}\ }\bibinfo
  {howpublished}
  {\url{www.engineeringtoolbox.com/helium-d_1418.html}}\BibitemShut {NoStop}%
\end{thebibliography}
\end{document}